\documentclass[10pt,journal,compsoc]{IEEEtran}

\usepackage{hyperref}
\usepackage[table]{xcolor}
\ifCLASSOPTIONcompsoc
  \usepackage[nocompress]{cite}
\else
  \usepackage{cite}
\fi

\usepackage{soul}
\usepackage{tabularx}
\usepackage{array}
\usepackage{graphicx}
\usepackage{amsmath}
\usepackage{amssymb}
\usepackage{multirow}
\usepackage{subcaption}
\usepackage{url}
\usepackage{floatrow}
\usepackage{pdflscape}
\usepackage{booktabs}
\usepackage{colortbl}
\usepackage{makecell}
\usepackage{tcolorbox}
\newcolumntype{a}{>{\columncolor{gray!35}}c}

\ifCLASSINFOpdf
\else
\fi

\begin{document}

\title{Agile Effort Estimation: Have We Solved the Problem Yet? Insights From A Replication Study}

\author{Vali~Tawosi,~
        Rebecca~Moussa,~
        Federica~Sarro~%
\IEEEcompsocitemizethanks{\IEEEcompsocthanksitem V. Tawosi, R. Moussa and F. Sarro are with the
Department of Computer Science, University College London, 
London, United Kingdom.\protect\\
E-mail: \{vali.tawosi, rebecca.moussa.18, f.sarro\}@ucl.ac.uk\\
For the purpose of open access, the author has applied a Creative Commons Attribution (CC BY) license to any Author Accepted Manuscript version arising.}%
}

\IEEEtitleabstractindextext{

\begin{abstract}
In the last decade, several studies have explored automated techniques to estimate the effort of agile software development.
We perform a close replication and extension of a seminal work proposing the use of Deep Learning for Agile Effort Estimation (namely Deep-SE), which has set the state-of-the-art since. 
Specifically, we replicate three of the original research questions aiming at investigating the effectiveness of Deep-SE for both within-project and cross-project effort estimation. We benchmark Deep-SE against three baselines (i.e., Random, Mean and Median effort estimators) and a previously proposed method to estimate agile software project development effort (dubbed TF/IDF-SVM), as done in the original study. 
To this end, we use the data from the original study and an additional dataset of 31,960 issues mined from TAWOS, as using more data allows us to strengthen the confidence in the results, and to further mitigate external validity threats.
The results of our replication show that Deep-SE outperforms the Median baseline estimator and TF/IDF-SVM in only very few cases with statistical significance (8/42 and 9/32 cases, respectively), thus confounding previous findings on the efficacy of Deep-SE. 
The two additional RQs revealed that neither augmenting the training set nor pre-training Deep-SE play lead to an improvement of its accuracy and convergence speed.
These results suggest that using semantic similarity is not enough to differentiate user stories with respect to their story points; thus, future work has yet to explore and find new techniques and features that obtain accurate agile software development estimates.
\end{abstract}


}

\maketitle
\IEEEdisplaynontitleabstractindextext
\IEEEpeerreviewmaketitle

\IEEEraisesectionheading{
\section{Introduction}
\label{sec:intro}}
\IEEEPARstart{I}n agile development, a software is realized through repeated iterations called sprints (usually 2--4 weeks), and is built in small incremental parts known as releases \cite{cohn2005agile}.
In each sprint, the development team designs, implements, tests and delivers a distinct product version or a working release. 
This allows adaptation to changing requirements at any point during the project life cycle.
Accordingly, the development team is required to complete a number of user-valued functionalities, called user stories, in every sprint ~\cite{trendowicz2014software, BeckFowler2000}.
User stories are specified in the form of one or two sentences in a user's everyday language.
Because of these small incremental iterations, it is essential for an agile team to focus on estimating the effort required to complete every single user story, rather than the whole project.
This allows them to prioritise the user stories, and to organise and manage a successful completion of sprints. The estimation at this granularity is referred to as task-level estimation herein.

Story Point (SP) is commonly used to measure the effort needed to implement a user story \cite{Usman2014,trendowicz2014software} and agile teams mainly rely on expert-based estimation \cite{cohn2005agile,deep2018}.
However, similar to traditional software project effort estimation~\cite{jorgensen2004review,sarro2020learning}, task-level effort estimation is not immune to the expert's subjective assessment \cite{Usman2014}. 
Subjective assessment may not only lead to inaccurate estimations but also, and more importantly to an agile team, it may introduce inconsistency in estimates throughout different sprints. Thus an intelligent task-level effort estimation technique seems exigent.

Abrahamsson et al. \cite{abrahamsson2011predicting} point out the three advantages of using an intelligent task-level effort estimation technique for Agile development.
First, an automated technique can use all the project information available from its inception and the history of the previous tasks to produce estimations for new ones.
Second, the decision of an automated technique is not influenced nor affected by any pressure from opinions of other individuals.
Third, an automated estimation is repeatable and predictable, thus it will always provide a consistent prediction.
One of the readily available information that intelligent task-level effort estimation techniques can use is the user story, which is composed by a title and a description of the issue usually and stored issue-tracking systems (see Table \ref{tbl.samplestories} for a few examples of user stories).

For more than a decade, researchers proposed techniques to predict the effort of a task in SP using features extracted from user stories \cite{abrahamsson2011predicting,porru2016,soares2018,scott2018,deep2018, tawosi2022investigating}.
In most cases, these studies urged that their model would be a decision support system for expert estimators in Agile teams.

In this paper, we explore the methods introduced to estimate story points for agile development at task-level and we perform a close replication and extension of the study which had introduced the current state-of-the-art, namely \mbox{Deep-SE}\footnote{Deep-SE has been published in 2019 in IEEE TSE and been cited 138 times (source Google Scholar) at the time of writing this paper.} (regarded as the ``original study'' in this paper) \cite{deep2018}.
In close replications, the conceptual, methodological, and substantive aspects of the original study are mostly kept invariant, and only minor variations are allowed to be introduced \cite{uncles2013designing}. 
This type of replication serves as an initial check to see if the findings of the original study hold, or can be generalized in slightly varied conditions like the validation against a new dataset.
If close replications confirm findings of an original study, an empirical generalization can be established \cite{juristo2010replication,uncles2013designing,santos2021comparing}.

\mbox{Deep-SE} is an end-to-end deep learning-based approach which accepts user story (i.e., issue title and description of a task recorded in Jira issue management system) as input and estimates story point as an output.
In this work, we replicate three out of six research questions investigated in the original study.\footnote{Two of the three research questions from the original study, which are not replicated herein, investigate the usefulness of Deep-SE's internal components by replacing them with simpler options.
Since the original study showed that the best results had been achieved with the components already used in Deep-SE, we do not replicate these two research question.
The third research question investigates the performance of Deep-SE in estimating adjusted story points.
They adjust the estimated story points of each issue using various information extracted from the issue reports after the issue was completed \cite{deep2018}.
We did not replicate this research question since the data used for this research question and the details to extract them were missing from the replication package provided by the original study \cite{deepseImpl}.
Moreover, the results in the original paper showed that using the adjusted story point benefited all the methods by reducing the range of the SP distribution for all projects.}
These research questions compare Deep-SE to three na\"ive baseline techniques, and to another task-level effort estimation technique introduced by Porru et al. \cite{porru2016} (we refer to it as ``TF/IDF-SVM'' herein), and they also investigate the suitability of Deep-SE for cross-project estimation.

Our results show that Deep-SE does not statistically significantly outperform the Median baseline or TF/IDF-SVM in at least half of the projects evaluated in the original study, which confound the results of the original study \cite{deep2018}. We also observe that Deep-SE's accuracy in cross-project estimation is not better than within-project estimation, which corroborates the original findings \cite{deep2018}.

In addition, we extend the study by evaluating the replicated research questions on an additional larger and more diverse dataset (consisting of 26 projects with 31,960 user stories in total), which we have curated and made publicly available (we refer to it as the Tawosi dataset). 
The extended evaluation of Deep-SE we carried out, using the Tawosi dataset, confirmed the findings of our replication, and strengthened the confidence in its results. The results confirmed the conclusion made in the original study only for one research question (cross-project estimation). As for the two other research questions (i.e., sanity check and comparison with TF/IDF-SVM), we found variations which led to a different conclusion than that reached by the original study. 
We discuss the possible origin of the variations in Section~\ref{sec:sp-replication discrepancies}.

We also pose two new research questions. 
The first one aims at investigating how using additional user stories from within-company and across different projects based on a realistic chronological order can affect Deep-SE's prediction performance. The second one explores the effectiveness of an expensive pre-training step used in the original study, on the accuracy of the estimation and on the speed of the convergence of Deep-SE. As stated by the original study, pre-training needs many hours to run and as much as 50,000 issues (without labels) per repository.\footnote{A repository in Jira is a collection of projects usually under development within a single company or a collection of inter-related organisations/teams which share resources and follow the same organisational regulations.}.

The answers to these research questions showed that augmenting the training set with issues from other projects has no appreciable effect on the prediction accuracy of Deep-SE. Also, pre-training the lower levels of Deep-SE does not only not significantly improve its accuracy, but it also does not have tangible impact on its convergence speed towards the best solution; thus, this step can be disregarded.

The rest of the paper is organised as follows. Section \ref{sec:related work} presents the related work, and Section \ref{sec:sp-replication background} briefly describes the structure of Deep-SE and TF/IDF-SVM. Section \ref{sec:sp-replicaiton study design} describes the design of our replication study in detail; including the research questions, the data and experimental method used to address them, and threats to validity. Section \ref{sec:sp-replication resulst} reports and discusses the results. Further insights on differences between the results of the original study and our replication are discussed in Section \ref{sec:sp-replication discrepancies}. Finally, Section \ref{sec:conclusion} concludes the paper with a plausible answer to the question ``\emph{Why couldn't Deep-SE outperform the baseline techniques for all the cases?}'' and suggestions for future work.

\section{Related Work}\label{sec:related work}

We summarise previous work proposing intelligent techniques to estimate story points in Table \ref{tbl.relatedwork}, and we describe each of them below in chronological order.

The first study, published in 2011, is the work of Abrahamsson et al. \cite{abrahamsson2011predicting}. They proposed to train a prediction model on 17 features extracted from user stories such as the priority and number of the characters in the user story, and 15 binary variables representing the occurrence of 15 keywords in the user stories.
These 15 keywords are the 15 most frequent terms in user stories. They used regression models, Neural Networks (NN), and Support Vector Machines (SVM) to build estimation models for two industrial case studies, one consisting of 1,325 user stories and the other of 13 user stories. The best results were obtained by SVM.

Five years later, Porru et al. \cite{porru2016} proposed to classify user stories into SP classes. Their approach uses features extracted from 4,908 user story descriptions recorded in Jira issue reports, collected from eight open-source projects. 
Specifically, they extract the type of the issue, the component(s) assigned to it, and the TF-IDF derived from the title and description of the issue. 
Their study confirmed Abrahamsson et al.'s findings that the user story and its length are useful predictors for story point estimation. 
Their results also suggest that more than 200 issues are needed for training the classifier to obtain a model with a satisfactory accuracy.

In 2018, Scott and Pfahl \cite{scott2018} used developer-related features alongside the features extracted from 4,142 user stories of eight open-source projects to estimate the story points using SVM. 
Developer-related features include the developer's reputation, workload, work capacity, and the number of comments. 
The results showed that using only developers’ features as input for the SVM estimator leads it to outperform Random Guessing, Mean, and Median baseline estimators.
This approach also outperformed two models using only features extracted from the text of the user story and using a combination of the developers’ and text features.

At the same time, Soares \cite{soares2018} proposed the use of different NLP techniques with auto-encoder neural networks to classify user stories based on the semantic differences in their title in order to estimate their effort in terms of SP. He used TF/IDF and document embedding with four variants of auto-encoders, and evaluated these models on 3,439 issue reports from six open-source projects. The results revealed no significant difference in the SP estimation accuracy of these approaches. Soares speculated that this might be due to the relative semantic simplicity of issue report titles. 

In 2019, Choetkiertikul et al. \cite{deep2018} proposed a new approach to SP estimation, based on the combination of two deep learning architectures to build an end-to-end prediction system, called Deep-SE. They used raw user story text as input to their system.
A word embedding was used to convert each word in a user story into a fixed-length vector, which is then fed to the deep learning architecture to map it to a space in which the semantically related stories are placed close to each other. In the final step, they used a regressor to map the deep representation into the SP estimate. They evaluated Deep-SE on 23,313 issues from 16 open-source projects and showed that it outperforms both baseline estimators and Porru et al.'s approach \cite{porru2016} based on Mean Absolute Error.

Subsequently, Abadeer and Sabetzadeh \cite{abadeer2021machine} evaluated the effectiveness of Deep-SE for SP prediction with a commercial dataset of 4,727 user stories collected from a healthcare data science company. They found that Deep-SE outperforms random guessing, mean and median baselines statistically significantly, however with a small effect size.

More recently, Fu and Tantithamthavorn \cite{fu2022gpt2sp} proposed GPT2SP, a Transformer-based deep learning model for SP estimation of user stories.
They evaluated GPT2SP on the dataset shared by  Choetkiertikul et al. \cite{deep2018} including 16 projects with a total of 23,313 issues. 
They investigated the performance of their model against Deep-SE for both within- and cross-project estimation scenarios, and evaluated the extent to which each components of GPT2SP contribute towards the accuracy of the SP estimates.
Their results show that GPT2SP outperforms Deep-SE
with a 6\%-47\% improvement over MAE for the within-project scenario and a 3\%-46\% improvement for the cross-project scenarios. 
However, when we attempted to use the GPT2SP source code made available by the authors, we found a bug in the computation of the Mean Absolute Error (MAE) which might have inflated the GPT2SP's accuracy reported in the original paper. Our proposed fixed has been accepted and merged into the repository at \url{https://github.com/awsm-research/gpt2sp/pull/2}. The results obtained based on this fix reveled that, in within-project scenario, GPT2SP outperforms the median baseline and Deep-SE in only six cases out of 16, where the difference is statistically significant in only three cases against median (two with negligible and one with small effect size), and two cases against Deep-SE (both with negligible effect size). \footnote{A technical report describing these results in more details can be found at \cite{tawosi2022gpt2sp}.}

\begin{table*}
\caption{Summary of the Related Work.}
\label{tbl.relatedwork}
\centering
\resizebox{\textwidth}{!}{
\begin{tabular}{ >{\raggedright\arraybackslash}p{0.2\linewidth} |  >{\raggedright\arraybackslash}p{0.2\linewidth}  >{\raggedright\arraybackslash}p{0.15\linewidth}  >{\raggedright\arraybackslash}p{0.1\linewidth} >{\raggedright\arraybackslash}p{0.25\linewidth}  >{\raggedright\arraybackslash}p{0.25\linewidth}} 
\toprule
Reference&Approach&Baseline&Measure&Data&Results\\
\midrule
Abrahamsson et al. \cite{abrahamsson2011predicting}, 2011, IEEE Int. Sym. on Empirical Software Engineering and Measurement&Regression-based ML (Regression, NN, SVR)&expert estimation&MMRE, MMER, PRED(25\%)&17 features extracted from 1,338 issue reports from two industrial projects&In one project the developers' subjective estimates work better than the proposed approach (SVR), in the other project the accuracy of SVR lies in the range of subjective estimates of developers\\
\midrule
Porru et al. \cite{porru2016}, 2016, Int. Conf. on Predictive Models and Data Analytics in Software Engineering&Classification-based ML (KNN, NB, DT, SVM), The SVM variant is referred to as ``TF/IDF-SVM'' in this paper&ZeroR (always selects as the outcome the most represented class)&MMRE&TF/IDF features and length of the title and description, and type and component of 4,908 issue reports from eight open-source projects & Estimating SP using an automated tool is feasible. SVM results in an MMRE between 0.16 and 0.61 after an initial training of more than 200 issues\\
\midrule
Scott and Pfahl \cite{scott2018}, 2018, Int. Conf. on Software and System Process&Classification-based ML (multiple two-class SVMs)&Mean, Median, and Random baselines&MAE and SA&TF/IDF features, length, and five developer-related features extracted from 4,142 issue reports from eight open-source projects&SVM classifier with only five developer-related features outperforms SVM with textual features or a combination of textual and developers' features\\
\midrule
Soares \cite{soares2018}, 2018, IEEE Int. Joint Conf. on Neural Networks&Classification-based ML (AutoEncoder NN)&TF/IDF with SVM&Precision, Recall and F-measure&Distributed memory paragraph vector and distributed bag-of-words paragraph vector extracted from the title of 3,439 issue reports from six open-source projects&Found no significant difference among the different variants of AutoEncoders used in the study regarding SP estimation accuracy\\
\midrule
Choetkiertikul et al. \cite{deep2018}, 2019, IEEE Transactions on Software Engineering&Regression-based deep-learning called Deep-SE (word embedding + LSTM + RHWN + regression)&Mean, Median, and Random baselines, and KNN&MAE, MdAE, and SA&Title and description of 23,313 issue reports from 16 open-source projects, and dataset used by Porru et al. \cite{porru2016}&Deep-SE outperforms the baselines and TF/IDF-SVM in all cases\\
\midrule
Abadeer and Sabetzadeh \cite{abadeer2021machine}, 2021, IEEE 29th International Requirements Engineering Conference Workshops (REW)&Deep-SE (word embedding + LSTM + RHWN + regression)&Mean, Median, and Random baselines&MAE, MdAE, and SA&Title and description of 4,727 issue reports from one industrial project&Deep-SE outperforms the baselines\\
\midrule
Fu and Tantithamthavorn \cite{fu2022gpt2sp}, 2022, IEEE Transactions on Software Engineering&Transformer-based deep-learning called GPT2SP &Mean and Median baselines and Deep-SE & MAE & Title and description of 23,313 issue reports from 16 open-source projects used by Choetkiertikul et al. \cite{deep2018}& GPT2SP outperforms Deep-SE and the baselines statistically significantly\\
\midrule
This Study&Deep-SE (word embedding + LSTM + RHWN + regression) and TF/IDF-SVM (SVM)&Mean, Median, and Random baselines&MAE, MdAE, and SA&Title, description, type and component of 31,960 issue reports from 26 open-source projects, in addition to the datasets used by Choetkiertikul et al. \cite{deep2018}&Deep-SE does not outperform the baselines and TF/IDF-SVM statistically significantly in all cases\\
\bottomrule
\end{tabular}
}
{\raggedright \scriptsize{Abbreviations: ML = Machine Learning / SVM = Support Vector Machine \cite{suthaharan2016support}/ SVR = Support Vector Regression \cite{corazza2013using} / NN = Neural Network \cite{specht1990probabilistic}/ KNN= K Nearest-Neighbour \cite{shepperd1997estimating}/ NB = Na\"ive Bayese \cite{rish2001empirical}/ DT = Decision Tree / LSTM = Long-Short Term Memory \cite{hochreiter1997long, gers1999learning} / RHWN = Recurrent Highway Network \cite{pham2016faster}/ MAE = Mean Absolute Error  / MdAE = Median Absolute Error / SA = Standard Accuracy \cite{shepperd2012evaluating, langdon2016exact} / MMRE = Mean Magnitude of Relative Error \cite{kitchenham2001accuracy}/ MMER = Mean Magnitude of Error Relative to Estimation (a.k.a estimation MMRE) \cite{kitchenham2001accuracy}/ PRED($l\%$)= Prediction at level $l$ \cite{kitchenham2001accuracy}} \par} 
\end{table*}

\section{The Deep-SE and TF/IDF-SVM Approaches for Agile Effort Estimation} \label{sec:sp-replication background}
This section provides an overview of the two story point effort estimation techniques used in this study, namely Deep-SE \cite{deep2018} and TF/IDF-SVM \cite{porru2016}, which make up the current state-of-the-art in Agile effort estimation.

Both Deep-SE \cite{deep2018} and TF/IDF-SVM \cite{porru2016} leverage the similarity between the target user story and previously estimated user stories to come up with an estimation for the target user story. TF/IDF-SVM relies on one of the simplest techniques for text similarity, namely the term frequency-inverse document frequency (TF/IDF) statistic. On the other hand, Deep-SE uses advanced techniques in deep-learning to exploit the semantic similarity between user stories.
When Choetkiertikul et al. \cite{deep2018} proposed Deep-SE, it was natural to compare it with TF/IDF-SVM.
In the following, we provide a brief description of each approach and refer the reader to the papers where they were proposed originally for a full description \cite{deep2018,porru2016}.

\subsection{Deep-SE} \label{sec:Deep-SE}
Choetkiertikul et al. \cite{deep2018} proposed Deep-SE as a deep learning model to estimate story point of a single software task, which is represented by an issue report, containing the title and description of the issue.

Their model is composed of four components: (1) Word Embedding, (2) Document representation using Long-Short Term Memory (LSTM) \cite{hochreiter1997long,gers1999learning}, (3) Deep representation using Recurrent Highway Network (RHWN) \cite{pham2016faster}, and (4) Differentiable Regression.

The first component converts each word in the title and description of issues (i.e., user story) into a fixed-length vector (i.e., word embedding). These word vectors serve as an input sequence to the LSTM layer, which computes a vector representation for the whole document. 
The document vector is then fed into the RHWN, which transforms the document vector multiple times, before outputting a final vector which represents the text. The vector serves as input to the regressor, which predicts the output story-point. The word embedding and LSTM layers are pre-trained without using SP values to come up with a proper parameter initialization for the main deep structure (referred to as the pre-trained language models) which helps achieve a faster convergence of the model.

Pre-training leverages two sources of information: the predictability of natural language, and the availability of free texts without labels (e.g., issue reports without story points). 
The first source comes from the property of languages that the next word can be predicted using previous words, thanks to grammars and common expressions \cite{deep2018}. As per Choetkiertikul et al.'s \cite{deep2018} suggestion, we used the pre-trained language models that they shared given that this step takes long to execute. 
The pre-trained language models are trained on a corpus of 50,000 issues selected from each story\footnote{No explanation is given in the original study on how the 50,000 issues are selected. We noticed that their pre-training dataset takes into use every issue available in a story until the number of issues reach 50,000.}.
The issues are not required to have SP estimation to be included in the pre-training stage and may also come from projects that are not considered for SP-estimation study. 

Moreover, Deep-SE dose not apply any pre-processing on the textual input (i.e., title and description) to remove non natural text like links and code snippets. Thus, we did not apply any pre-processing in order to perform a close replication.

Furthermore, Deep-SE aims at estimating a value closest to the target SP value. To this end, it uses a regressor in its last layer and, thus, the estimate could be a real number rather than an integer.
Although the use of the Fibonacci series, the T-Shirt size, or the Planning Poker card set\footnote{Planing Poker card set includes the following numbers: 0, 0.5, 1, 2, 3, 5, 8, 13, 20, 40, 100, and $\infty$.} for SP scales is usually recommended and applied by expert estimation and planning poker practitioners \cite{cohn2005agile, ziauddin2012effort}, Deep-SE's estimations would not follow a Fibonacci scale, and the authors did not round the predicted SPs to the nearest SP integer value on the Fibonacci scale.

\subsection{TF/IDF-SVM} 
\label{sec:porru's method}
Porru et al. \cite{porru2016} treat SP estimation as a classification problem.
Their idea is to use the title and description of issues (i.e., user stories) and issue related data available at the estimation time (i.e., issue type and component), to build a machine learning classifier for estimating the story points required to address an issue.
They extract TF/IDF statistics and user story length from issue reports, and use issue type and the component of the issue to build a dataset in which rows represent issues and columns are the TF/IDF values alongside the binary values representing the association of an issue to a type or component. Specifically, they concatenate the title and description of the issue reports and call it context.
They separate the code snippets (if any) from natural language text in the context and analyse two chunks separately to extract TF/IDF features.
They concatenate three feature sets, two extracted from text and code chunks, and one extracted from issue type and components, before reducing the dimension by feature selection methods. The selected features are then fed to a Support Vector Classifier to classify issues into SP classes. 

Given that the TF/IDF-SVM method treats the task-level effort estimation problem as a multi-classification one (i.e., each class label is a value in the Fibonacci scale), the number of classes must be determined a-priory. Thus, its adoption might not be directly applicable to projects that do not follow the Fibonacci scale \cite{deep2018}.

\section{Empirical Study Design} 
\label{sec:sp-replicaiton study design}
In this section, we provide a detailed description of the research questions investigated for the replication and extension of the original study \cite{deep2018}, together with the data,  benchmarks, and  evaluation methods used to answer these questions.
To carry out both the replication and extended study, we use the implementation and configuration made publicly available by 
Choetkiertikul et al. \cite{deep2018} for both \mbox{Deep-SE} and \mbox{TF/IDF-SVM} \cite{deepseImpl}, subject to some alterations needed to fix some errors.\footnote{Since Porru et al. \cite{porru2016} did not provide any public implementation for TF/IDF-SVM, we use the one provided by Choetkiertikul et al. \cite{deepseImpl} Moreover, the link to the replication package provided in Porru et al. \cite{porru2016} is no longer maintained by the authors and when contacted the authors were not able to provide us with more data nor tools.}.
To allow the replication of our work, our scripts and data are publicly available \cite{replicationPackage}.

\subsection{Research Questions} \label{sec:sp-replication rqs}
The first research question asked by Choetkiertikul et al. \cite{deep2018} investigates  whether Deep-SE is a suitable method to estimate story points. To this end they assess whether it is able to outperform simpler baseline estimators for within-project estimation (i.e., building a prediction model by using past issues of a given project, and using such a model to predict the story points for new issues within the same project).

Specifically, they compare Deep-SE with the Mean and Median baselines and they also report the SA values for all three techniques. SA is a measure of their accuracy performance against Random Guessing (see Sections \ref{baselines} and \ref{sec:evaluation} for the definition of the benchmarks and the evaluation method).
Therefore, in our replication we investigate the same research question:

\textbf{RQ1. Sanity Check:} \textit{Is Deep-SE suitable for story point estimation?}

We evaluate this research question through two sub-questions. 
We first replicate the original study by evaluating Deep-SE on the same data used by Choetkiertikul et al. \cite{deep2018}:  \textbf{RQ1.1. Sanity Check - Replication}, where  we compare Deep-SE's accuracy with baseline estimators and the results reported in the original study.

Second, we extend RQ1.1 by evaluating Deep-SE on a larger dataset comprising 26 projects (see Section~\ref{sec:data} for detail) to further strengthen the confidence in the results of RQ1: \textbf{RQ1.2. Sanity Check - Extension.} 

If we find that Deep-SE outperforms the baseline techniques for both RQ1.1 and RQ1.2, then we can confirm the conclusion made in the original study.

Choetkiertikul et al. \cite{deep2018} also compared Deep-SE to previous state-of-the-art for within-project estimation proposed by Porru et al. \cite{porru2016} (i.e., \mbox{TF/IDF-SVM}). 
This motivates our second research question:

\textbf{RQ2. Deep-SE vs. TF/IDF-SVM:} \textit{How does Deep-SE perform against TF/IDF-SVM in story point estimation?}

Similarly to RQ1, we investigate two sub-questions for RQ2 by replicating the same experiment with the same data used in the original study (\textbf{RQ2.1. Deep-SE vs. TF/IDF-SVM - Replication}) and by extending it and experimenting with a larger dataset (\textbf{RQ2.2. Deep-SE vs. TF/IDF-SVM - Extension}).
Deep-SE outperforming TF/IDF-SVM for both RQ2.1 and RQ2.2 will further confirm the conclusion made in the original study. 

Another important question tackled by the original study is the performance of Deep-SE for cross-project estimation, i.e.,  building a prediction model by using issues from another project, and using such a model to predict the story points for the issues of a new target project, for which there are no issues to train an accurate model \cite{porru2016,deep2018}.
Cross-project estimation is generally deemed as a more difficult task than within-project estimation, since the training and target projects might be heterogeneous \cite{MendesKMFS14,minku2015make,ferrucci2012web}. Thus, our third research question assesses the ability of Deep-SE in cross-project estimation, as posed in the original study:

\textbf{RQ3. Cross-project Estimation.} \textit{Is Deep-SE suitable for cross-project estimation?}

We tackle this question by replicating the two experiments done by Choetkiertikul et al. \cite{deep2018} in our first sub-question \textbf{RQ3.1. Cross-project Estimation - Replication}. Specifically, in the first experiment to train Deep-SE we use a project belonging to the same repository of the target project (i.e., \textit{cross-project within-repository estimation}). In the second experiment, we use as training data the issues from a project belonging to repository different from that of the target project (i.e., \textit{cross-project cross-repository estimation}).
We replicate this research question with the same set of the source and target projects and compare the results of Deep-SE to the Mean and Median baselines, as done in the original study.
We observe that Choetkiertikul et al. \cite{deep2018} did not consider the temporal ordering of the issues' creation time in collecting issues for their training and testing sets for this research question. 
However, previous work on software engineering prediction models showed that neglecting contextual details (like the temporal ordering of the data) is a potential threat to the conclusion stability of the study \cite{bangash2020time,JimenezRPSTH19,falessi2020need}. 
In a real-world scenario, one would be able to use only the issues from the past projects to train the model. 
Thus, all the issues in the training (and validation) set should precede, in creation time, the earliest created issue from the test set.
Surprisingly, for 11 out of 16 experiments, more than 97\% of the issues used to train the models in RQ3.1 were created after the start date of the target project, thus making the use of Deep-SE in practice infeasible.\footnote{Among the five remaining projects, for two of them, 75\% and 37\% of the issues were created after the start date of the target projects, and for three (in which MULESOFT was used as the source or target project) we could not determine the percentage as the start/end time of the MULESOFT project is not known as this project repository is no longer accessible.}
Therefore, we extend the original analysis in order to assess Deep-SE's performance in a realistic scenario, which takes into account the chronological order of the data used for cross-project estimation. This motivates, our second sub-question: \textbf{RQ3.2. Cross-project Estimation - Extension}, where we only use the issues created before the start date of the target project as the cross-project training data for Deep-SE. We compare the results of Deep-SE with Mean and Median baseline techniques.

Our next research question, focuses on another practical usage scenario: The case where an engineer needs to estimate the effort for a target project that is new and for which there are not enough story issues realised yet; thus there is no enough data from the project itself to train an accurate deep-learning-based prediction model. As a result, the data would need to be augmented with the use of external sources.
This is a very common case for deep learning models, which need an abundance of training data to perform well, and previous work investigates how to augment the training set in different application domains \cite{taylor2017improving,mikolajczyk2018data}.
Our research question aims at analysing if augmenting the training set (composed of a few instances from target project) with a larger number of issues from other projects, helps Deep-SE produce more accurate estimates:

\textbf{RQ4. Augmented Training.} \textit{How does Deep-SE perform when using a training set augmented with other projects issues?}

To answer RQ4, we use the same data splits used for RQ1.2, but we add to the training set historical issues available from all the other projects belonging to the same repository of the target project (i.e., all projects are within a same organisation), yet respecting the chronological order of the issues.
In particular, we include in the training set only those issues from other projects within a same  repository that are created before the earliest created issue in the validation set.
Therefore, we augment the target project training data with every issue that a company already had in their repository at the time of predicting the story points for new issues of a target project.
Using more instances in the training data changes its original distribution, however, the same data is used to train all approaches and no transformation is applied to the test set. This allows for a fair and correct empirical analysis answering RQ4.

Our fifth and last research question, investigates a technical aspect of the deep learning approach proposed by Choetkiertikul et al. \cite{deep2018}: The utility of pre-training  for Deep-SE.
Pre-training is described as a way to come up with a good parameter initialization for Deep-SE without using the labels (i.e., story points). Choetkiertikul et al. \cite{deep2018} pre-train the lower layers of Deep-SE (i.e., embedding and LSTM layers), which operate at word and sentence level. 
In the absence of pre-training, the parameters are typically initialized randomly, but Choetkiertikul et al. \cite{deep2018} stated that a good initialization (through pre-training of embedding and LSTM layers) allows a faster convergence towards good solutions.
However, their study did not compare the performance of Deep-SE with and without using such a pre-training step. In this work, we investigate the actual benefit of using such a pre-training step since pre-training is very expensive both in terms of running time and amount of data required (Choetkiertikul et al. reports that their pre-training step needs around 50,000 issues to work properly and took 46 hours and 48 minuets for nine repositories to run\cite{deep2018}).
Given these limitations, if the use of pre-training does not improve the estimation performance (accuracy-wise and execution time-wise), its usage would not be advisable. This motivates our next research question:
 
\textbf{RQ5. Pre-Training Effectiveness.} \textit{Does pre-training Deep-SE's lower levels improve its accuracy and/or convergence speed?}

To answer this question, we compare Deep-SE with random weight initialization to Deep-SE initialized with weights tuned by pre-training (i.e., the pre-training step used by  Choetkiertikul et al. \cite{deep2018}).

The comparison is performed on estimation accuracy, running time, and the number of epochs each of them performs before converging to an optimal solution.
We perform this analysis on both, the Choet and Tawosi datasets.

\subsection{Datasets} \label{sec:data}
To execute a close replication of the research questions to that of the the original study, we adopted the datasets made publicly available by Choetkiertikul et al. \cite{deep2018}. Moreover, we also gather our own dataset in order to strengthen our confidence in the results and further mitigate the threat to the external validity of the study. We explain each of these datasets in the following, while a summary of the descriptive statistics of all three datasets is in Table \ref{tbl.ourds-porru-stat}.

\subsubsection{The Choetkiertikul Dataset}
Choetkiertikul et al. \cite{deep2018} mined 16 open-source projects from nine different repositories (namely Apache, Appcelerator, DuraSpace, Atlassian, Moodle, Lsstcorp, MuleSoft, Spring, and Talendforge) and gathered a total number of 23,313 issues (i.e., user stories) with recorded story points, after filtering out all those issues that had assigned a story point of zero, or negative, or an unrealistically large value (i.e., greater than 100).
Choetkiertikul et al. collected their dataset on August 8, 2016, and made it publicly available \cite{deepseImpl}. We refer to this dataset as the Choet dataset and we use it to answer RQs 1.1, 3.1, and 5.

\subsubsection{The Porru Dataset}
\label{porrudata}
Choetkiertikul et al. \cite{deep2018} also collected eight open-source projects stored in six open-source repositories, aiming at  benchmarking Deep-SE against Porru's TF/IDF-SVM approach using a common dataset.\footnote{The dataset used in Porru et al. \cite{porru2016} is not publicly available, so Choetkiertikul et al. \cite{deep2018} re-collected it by closely following the approach described in their paper \cite{porru2016}. This dataset slightly differs from the  the original dataset \cite{porru2016} in the number of issues.}

In total, the Porru dataset, as collected by Choetkiertikul et al. \cite{deep2018,deepseImpl}, contains 4,904 issues. Among these eight projects, six are common with the Choet dataset (i.e., TIMOB, TISTUD, APSTUD, MESOS, MULE, and XD), although they contain a different subset of issues as Porru et al. applied a set of more restrictive filtering criteria than those used by Choetkiertikul et al. in building the Choet dataset \cite{deep2018}. These criteria are as follows:
  (1) Story points must have been assigned once and never updated afterwards. Updated story points may mean that the information provided in the issue report had misleading information, and they do not want to confuse the classifier with noisy input. 
  (2) The issue must be addressed. Issues not addressed are likely to be unstable; hence they might also confuse the classifier.
  (3) Once the story points are assigned, the informative fields of the issue (i) must be already set and (ii) their value must not have been changed afterwards. 
  They define informative fields used as dependent variables for classification, including Issue Type, Description, Summary, and Component(s). 
  (4) The values in the story points field must correspond to one of those included in the Planning Poker cards set. 

Choetkiertikul et al. made their version of the Porru dataset publicly available \cite{deepseImpl}. We refer to this dataset as the Porru dataset.  We used this dataset to answer RQ2.1.

\subsubsection{The Tawosi Dataset}
In our empirical study, we also used issues extracted from the TAWOS dataset \cite{tawosi2022versatile}, which contains over 500,000 issues, coming from 44 open-source projects found in 13 Jira repositories.
Specifically, we sampled from TAWOS all those issues having a SP value specified, and then filtered out all those that do not meet the criteria recommended by Porru et al. \cite{porru2016} (see Section \ref{porrudata}) as they are more restrictive than those used by Choetkiertikul et al. \cite{deep2018}, and could be more effective in removing noisy data points \cite{porru2016}. 
Our sample resulted a total of 31,960 issues coming from 26 different projects (each having more than 200 issues). We will refer to this sample as the \textit{Tawosi} dataset, herein. The \textit{Tawosi} data is available in our on-line replication package \cite{replicationPackage}, and we used it in our empirical study to answer RQs 1.2, 2.2, 3.2, and 4.

The 26 projects in the Tawosi dataset belong to 13 different Jira repositories. Eleven of these 13 repositories were used by either Choetkiertikul et al. \cite{deep2018} or Porru et al. \cite{porru2016} and other previous work \cite{soares2018,scott2018}, and the two additional ones, Hyperledger and MongoDB, were used in related work \cite{choetkiertikul2017predicting}.

Twelve of the 26 projects in the Tawosi dataset are in common with the Choet dataset, however they have a different number of issues, as we collected the data more recently and applied Porru's filtration criteria.\footnote{Four projects in our dataset have more issues than the same projects in the Choet dataset (4,652 additional issues in total for the TIMOB, DM, MDL, and MULE projects). This is due to the new issues created in these projects since Choet collected issues for their dataset. On the other hand, eight projects in our dataset have smaller number of issues than the same projects in the Choet dataset (4,424 fewer issues in total for the MESOS, TISTUD, APSTUD, CLOV, DURACLOUD, XD, TDQ, and TESB projects). This is due to the more restrictive issue filtration policy that we followed in this study.}
In addition, the Tawosi dataset includes 14 more projects (namely, ALOY, CLI, DAEMON, TIDOC, CONFCLOUD, CONFSERVER, DNN, FAB, STL, COMPASS, SERVER, EVG, NEXUS, and TDP) with a total of 10,506 issues. Four projects in the Choet dataset (with a total of 2,087 issues) are not included in the Tawosi dataset: Three of them (i.e., USERGRID, BAM, and JRESERVER) were left with less than 200 issues after applying the Porru's filter, thus removed from the Tawosi dataset; while the MULESTUDIO project was no longer available on Jira, when the TAWOS dataset was collected.

The Tawosi dataset contains all eight projects as included in Porru's dataset, with 11,029 more issues. Moreover, the Tawosi dataset includes 18 additional projects with respect to the Porru's dataset, with a total number of 16,027 additional issues.

\bgroup
\rowcolors{5}{white}{gray!25}
\begin{table*}
\caption{Descriptive statistics of the three datasets used in this study.}
\label{tbl.ourds-porru-stat}
\centering
\resizebox{\textwidth}{!}{
\begin{tabular}{l | l l | r | r r r r r | r | r r r r r | r | r r r r r }
\toprule
 \multirow{3}{*}{Repository}  & \multirow{3}{*}{Project (Abbreviation used in \cite{deep2018})}   & \multirow{3}{*}{Key} & \multicolumn{6}{l}{Tawosi dataset} & \multicolumn{6}{|l}{Choet dataset \cite{deep2018}} & \multicolumn{6}{|l}{Porru dataset \cite{deep2018}} \\\cmidrule{4-21}
 &  &  &   & \multicolumn{5}{c|}{Story Point} &   & \multicolumn{5}{c|}{Story Point} &   & \multicolumn{5}{c}{Story Point} \\\cmidrule{5-9}\cmidrule{11-15}\cmidrule{17-21}
 &  &  & \multirow{-2}{*}{\#Issues} & Min & Max & Mean  & Median  & Std & \multirow{-2}{*}{\#Issues} & Min & Max & Mean  & Median  & Std & \multirow{- 2}{*}{\#Issues}  & Min & Max & Mean  & Median  & Std \\ 
 \midrule
 \hline
\cellcolor{white}&Mesos (ME) &MESOS&1,513&0&13&3.15&3&2.14&1,680&1&40&3.09&3&2.42&354&0&13&2.8&2&2.15\\
\multirow{-2}{*}{Apache}&Usergrid (UG) &USERGRID&&&&&&&482&1&8&2.85&3&1.40&&&&&&\\
\hline
\hline
\cellcolor{white}&Alloy&ALOY&241&0&13&3.71&3&2.32&&&&&&&&&&&&\\
&Appcelerator Studio (AS)&TISTUD&2,794&0&40&5.48&5&3.23&2,919&1&40&5.64&5&3.33&1,172&1&144&5.53&5&4.79\\
\cellcolor{white}&Aptana Studio (AP)&APSTUD&476&0&100&7.93&8&7.19&829&1&40&8.02&8&5.95&236&0&100&7.76&8&7.58\\
&Command-Line Interface&CLI&293&0&13&3.18&3&2.30&&&&&&&&&&&&\\
\cellcolor{white}&Daemon&DAEMON&205&1&13&5.58&5&3.76&&&&&&&&&&&&\\
&Documentation&TIDOC&1,005&0&40&3.58&2&3.68&&&&&&&&&&&&\\
\cellcolor{white}\multirow{-7}{*}{Appcelerator}&Titanium (TI)&TIMOB&3,915&0&20&4.68&5&3.32&2,251&1&34&6.32&5&5.10&600&0&6,765&16.89&5&276.19\\
\hline
\hline
&Bamboo (BB)&BAM&&&&&&&521&1&20&2.42&2&2.14&&&&&&\\
\cellcolor{white}&Clover (CV)&CLOV&336&0&100&5.33&2&11.03&384&1&40&4.59&2&6.55&&&&&&\\
&Confluence Cloud&CONFCLOUD&234&0&13&2.91&2&2.24&&&&&&&&&&&&\\
\cellcolor{white}&Confluence Server and Data Center&CONFSERVER&456&0&13&3.12&3&1.93&&&&&&&&&&&&\\
\multirow{-5}{*}{Atlassian}&Jira Server and Data Center (JI)&JRESERVER&&&&&&&352&1&20&4.43&3&3.51&&&&&&\\
\hline
\hline
\cellcolor{white}DNNSoftware&DNN&DNN&2,064&0&100&2.05&2&2.56&&&&&&&866&0&13&1.91&2&1.25\\
\hline
\hline
DuraSpace&Duracloud (DC)&DURACLOUD&310&0&20&1.72&1&1.70&666&1&16&2.13&1&2.03&&&&&&\\
\hline
\hline
\cellcolor{white}&Fabric&FAB&303&0&40&2.69&2&3.20&&&&&&&&&&&&\\
\multirow{-2}{*}{Hyperledger}&Sawtooth&STL&206&0&5&2.09&2&1.19&&&&&&&&&&&&\\
\hline
\hline
\cellcolor{white}Lsstcorp&Data Management (DM)&DM&5,381&0&100&3.05&2&6.87&4,667&1&100&9.57&4&16.60&&&&&&\\
\hline
\hline
&Compass&COMPASS&260&1&8&3.55&3&1.85&&&&&&&&&&&&\\
\cellcolor{white}&Core Server&SERVER&519&0&20&2.58&2&2.40&&&&&&&&&&&&\\
\multirow{-3}{*}{MongoDB}&Evergreen&EVG&2,824&0&8&1.43&1&0.86&&&&&&&&&&&&\\
\hline
\hline
\cellcolor{white}Moodle &Moodle (MD)&MDL&1,394&0&100&11.80&5&18.81&1,166&1&100&15.54&8&21.65&&&&&&\\
\hline
\hline
&Mule (MU)&MULE&2,935&0&13&3.88&3&3.46&889&1&21&4.90&5&3.61&810&0&13&3.42&3&3.43\\
\cellcolor{white}\multirow{-2}{*}{MuleSoft}&Mule Studio (MS)&MULESTUDIO&&&&&&&732&1&34&6.40&5&5.39&&&&&&\\
\hline
\hline
Sonatype&Sonatype’s Nexus&NEXUS&1,425&0&40&1.70&1&1.82&&&&&&&420&0&8&1.13&1&0.86\\
\hline
\hline
\cellcolor{white}Spring&SpringXD (XD)&XD&811&0&20&3.16&3&2.56&3,526&1&40&3.70&3&3.23&446&0&8&2.77&2&2.08\\
\hline
\hline
&Talend Data Preparation&TDP&471&0&13&2.31&2&1.84&&&&&&&&&&&&\\
\cellcolor{white}&Talend Data Quality (TD)&TDQ&859&0&40&6.01&5&4.66&1,381&1&40&5.92&5&5.19&&&&&&\\
\multirow{-3}{*}{Talendforge}&Talend ESB (TE)&TESB&730&0&13&2.13&2&1.45&868&1&13&2.16&2&1.50&&&&&&\\
\hline
\midrule
\rowcolor{white}
Total&&&31,960&&&&&&23,313&&&&&&4,904&&&&&\\
\bottomrule
\end{tabular}
}
\end{table*}
\egroup

\subsection{Validation Approach}
\label{sec:validation}
Similarly to the original study, to perform within project estimation (i.e., for RQs 1, 2, and 5), for all datasets, we sorted the issues in each project in ascending order of their creation time and used the 60\% oldest issues as the training set, 20\% as the validation set and the newest 20\% as the testing set, as has been done in the previous studies. 
This is because, in a real-world scenario, estimation for a new issue is made by using knowledge from the estimations of the past issues.
In particular, for RQs 1.1, 2.1, and 5, which are close replications, we used the same data splits used by Choetkiertikul et al. \cite{deep2018} to avoid any possible bias in the data that might be introduced by using a different sampling.

For RQ3 (cross-project estimation), the original study does not clarify how the data was split. Thus, we apply the same rate they used for RQ1.1 and divided the source project into three-forth (75\%) train and one-fourth (25\%) validation sets (following the 60\% train-20\% validation split rate) and used all issues of the target project as test set.

Finally, for RQ4 (augmented training set), we use the same train-validation-test as RQ1.2, but we augmented the training set with additional issues from different projects within a same repository of the target one.

\subsection{Benchmarks} \label{baselines}
As done in the original study, we use three commonly used baseline estimation techniques, i.e., Random Guessing, Mean, and Median estimators to benchmark Deep-SE.

\textbf{Random Guessing (RG)} is a na\"ive method that simply assigns story point of a randomly selected issue to the target issue \cite{shepperd2012evaluating}.
More formally, Random Guessing predicts a story point value $y$ for the target case $issue_t$ by randomly sampling (with equal probability) all the remaining $n-1$ cases and taking $y = r$; where $r$ is the story point of the randomly drawn $issue_r$ from $1...n~|~issue_r \neq issue_t$ \cite{shepperd2012evaluating}. 
This method does not need any parameter estimation and any prediction system is expected to outperform it over time, otherwise, the prediction system would not be using any information about the target case. The results for RG are in the on-line appendix \cite{replicationPackage}.

\textbf{Mean and Median Effort} are two baseline benchmarks commonly used for effort estimation techniques \cite{mittas2015framework, whigham2015baseline, sarro2016multi}. Specifically, the mean or median story point of the past issues is used as the predicted story point for a new issue.

\subsection{Evaluation Measures and Statistical Analysis} \label{sec:evaluation}
Several measures have been used in the software effort estimation literature to measure the accuracy of the estimation models.
These measures are generally built upon the error (or absolute error) between the predicted value and the actual value (i.e., $|Actual.value - Predicted.value|$).

Similarly to the original study, we discuss all results of our study based on the MAE measure, while we report the MdAE and SA values in our on-line appendix for completeness \cite{replicationPackage}.

These measures (defined in Equations \ref{eq:mae}, \ref{eq:mdae} and \ref{eq:sa}) are standardised measures which are not biased towards under- or over-estimates and which have been recommended in previous work \cite{shepperd2012evaluating, langdon2016exact, sarro2016multi}.

Across $n$ issues, the MAE and MdAE are computed as follows: 

\begin{equation}\label{eq:mae}
MAE = \frac{1}{n}\sum_{i=1}^n |actual_i - predicted_i|
\end{equation}  

\begin{equation}\label{eq:mdae}
MdAE = Median{_{i=1}^n \Big\{|actual_i - predicted_i|}\Big\}
\end{equation}  

\noindent where $actual_i$ is the actual effort, $predicted_i$ is the predicted effort and $n$ is the number of issues in a given project.

$SA$ is recommended as a standard measure to compare multiple prediction models against each other \cite{shepperd2012evaluating}. It is based on MAE and is defined as follows:

\begin{equation} \label{eq:sa}
SA = \left( 1 - \frac{MAE_{p_i}}{MAE_{p_0}} \right) \times 100
\end{equation}

\noindent where $MAE_{p_i}$ is the $MAE$ of the approach $p_i$ being evaluated and $MAE_{p_0}$ is the $MAE$ of a large number (usually $1,000$ runs) of \textit{random guesses}.

For a prediction model $p_i$ which outperforms random guessing in terms of accuracy, $SA$ will produce a number in the range $[0, 1]$.
An $SA$ value closer to zero means that the predictor $p_i$ is performing just a little better than random guessing \cite{sarro2016multi,shepperd2012evaluating}.
For a prediction model which is outperformed by random guessing $SA$ will produce a negative value.
For a high-performance prediction model $MAE$ and $MdAE$ should be lower and $SA$ should be higher than the competitors.

To check if the difference in the results achieved by two methods is statistically significant, we performed the Wilcoxon Ranked-Sum test (a.k.a. Mann–Whitney U test) on the distribution of the absolute errors produced by the methods under investigation. Specifically, we used a one-sided Wilcoxon test with a confidence limit of $\alpha$ = $0.05$  to check the following Null Hypothesis:
\noindent{\textbf{$H_0$}:}
\textit{The distribution of absolute errors produced by the two prediction models $P_i$ and $P_j$ are not different.}
If the test rejects the Null Hypothesis, the alternative hypothesis would be accepted:
\noindent{\textbf{$H_1$}: \textit{The absolute errors produced by the $P_i$ are different from those provided by the $P_j$.}}
Specifically, we performed a one-sided test since we are interested in knowing if a given model (i.e., Deep-SE) would commit a smaller estimation errors than another model. In such a case, the one-sided p-value interpretation would be straightforward.
To mitigate the risk of incorrectly rejecting the Null Hypothesis (i.e., Type I error) \cite{sarro2020learning}, we also analyse how the results would be when the Bonferroni correction is applied to cater for multiple hypothesis testing  (i.e., the confidence limit is set as $\alpha/K$, where $K$ is the number of hypotheses).
Therefore, herein we report the original p-value results of the Wilcoxon test (see Tables \ref{tbl.wilcox.replication}, \ref{tbl.wilcox.porru.sp.myds}, \ref{tbl.wilcox.porru}, \ref{tbl.cross-project}, and \ref{tbl.within-repo-rq3.2}). While, we analyse and discuss the results both using the confidence limit of $\alpha$ = $0.05$ and the Bonferroni corrected one ($\alpha$ = $0.025$), where applicable.

We also use a standardised non-parametric effect size measure (i.e., the Vargha Delaney's $\hat{A}_{12}$ statistic) to assess the practical magnitude of the difference between two methods, as recommended in previous work \cite{sarro2016multi,arcuri2014hitchhiker,sarro2020learning}.
For two algorithms $A$ and $B$, the $\hat{A}_{12}$ measures the probability of $A$ performing better than $B$ with respect to a performance measure.
$\hat{A}_{12}$ is computed using Equation~(\ref{equation:A12}), where $R_1$ is the rank sum of the first data group being compared, and $m$ and $n$ are the number of observations in the first and second data sample, respectively.

\begin{equation}\label{equation:A12}
\hat{A}_{12} = \frac{(\frac{R_1}{m} - \frac{m + 1}{2})}{n}
\end{equation} 

Based on Equation~(\ref{equation:A12}), if two algorithms are equally good, $\hat{A}_{12} = 0.5$. 
Respectively, $\hat{A}_{12}$ higher than $0.5$ signifies that the first algorithm is more likely to produce better predictions.
The effect size is considered negligible for \mbox{$\hat{A}_{12} < 0.6$} (represented by an `n'), small (s) for \mbox{$0.6 \leq \hat{A}_{12} < 0.7$}, medium (m) for \mbox{$0.7 \leq \hat{A}_{12} < 0.8$}, and large (l) for \mbox{$\hat{A}_{12} \geq 0.8$}, although these thresholds are not definitive \cite{sarro2016multi}. We do not transform the $\hat{A}_{12}$ as we are interested in \textit{any} improvement achieved by the methods \cite{neumann2015transformed, sarro2016multi}.

To perform the above analyses, we used the Wilcoxon Rank-Sum test and Vargha Delaney's $\hat{A}_{12}$ effect size available from the {\tt stats} library in {\tt R} v. 4.0.1 \cite{r}. 

\subsection{Threats to Validity} 
\label{sec:thread to validity}

Our study is a close replication and extension of a previous work on task-level software effort estimation\cite{deep2018}. Thus, it shares some of the threats to the validity of the original study, but also has implemented further mitigation.
We followed best practice for designing the replication and reporting finding in a sound way \cite{shull2002replicating,juristo2010replication,uncles2013designing,santos2021comparing,carver2010towards}.

In the original study, the authors expressed their concern about basing their ground truth on the most likely biased human-estimated story points \cite{tawosi2022relationship}. However, they argued that Deep-SE is trained to imitate human beings with respect to estimations by reproducing an estimate that human engineers would be able to derive. For this aim, story points sufficiently serve the purpose. Moreover, whenever an unbiased ground truth is available, Deep-SE can be trained on the new target variable.

To minimize threats to conclusion validity, the original study, and similarly our study, used unbiased accuracy performance measures and applied statistical tests by checking all the required assumptions. Besides, in our study, we checked by performing peer-code-review that the implementation of all estimation approaches and the computation of the accuracy measures adhere to their original definitions.

To mitigate the external validity, like the original study, we collected data from real-world open-source projects to evaluate the methods.
Although these projects differ greatly in size, complexity, and developers community, we cannot claim that they are representative of all kinds of software projects.
Specially, there are differences between open-source and commercial software projects. 
A key difference, that may affect the estimation of story points, is the behaviour of contributors, developers, and the project’s stakeholders.
It is expected that in an industrial setting for a commercial software project, the user stories are written in a more cohesive and disciplined manner, thus, providing more useful information and containing less noise.
Hence, further investigation for commercial projects from industrial software companies is needed to strengthen the conclusions made in this study.

Santos et al. \cite{santos2021comparing} showed that sampling error and (un)intentional contextual modifications in experimental settings can produce different results, leading to an unsuccessful reproduction of results. To avoid such errors and modifications, we made sure that for the replicated RQs we use the same data and train-validation-test as the ones used in the original study.
We also made sure that all the projects have more than 200 observations, thus mitigating the threat of obtaining unstable p-values and effect sizes due to undersampling \cite{santos2021comparing}.
To ensure that we keep the experimental setting consistent with the original study, we followed the procedures outlined in their paper.
Furthermore, for both Deep-SE and TF/IDF-SVM, we used the implementation provided by Choetkiertikul et al. \cite{deep2018}.
The only changes we made are to amend and fix errors found in the implementation of the baseline techniques and evaluation metrics according to the definition recommended in the literature (see Section \ref{sec:sp-replication discrepancies}).

Our scripts and dataset are publicly available  \cite{replicationPackage}. 

\section{Results}\label{sec:sp-replication resulst}
This section presents the results and discusses the answers to the research questions.

\subsection{RQ1. Sanity Check}

\subsubsection{RQ1.1. - Replication}
\label{RQ1.1replication}
\textbf{Results:} In Table ~\ref{tbl.replication}, under the column ``Rep'', we report the results obtained by Deep-SE and the baseline estimators (i.e., Mean and Median) on the Choet dataset in our replication study. Below we analyse and discuss the results based on the MAE measure.

We can observe that Deep-SE generally outperforms the Mean baseline estimator obtaining lower MAE values on 14 out of the 16 projects (i.e., 88\%) investigated in our replication study. The differences on 11 of them (i.e., 69\%) is statistically significant with two projects having a medium effect size, six having small, and three having negligible ones as shown in Table~\ref{tbl.wilcox.replication}. 
However, when Deep-SE is compared to the Median estimator, results reveal that the former obtains lower MAE values on 8 out of the 16 projects under study (i.e., 50\%) with the differences being statistically significant in only 4 cases (25\%) with the effect size being small in one case and negligible in the remaining three cases. All these cases remain statistically significant when considering the Bonferroni correction.

\textbf{Discussion:} These results are not in complete accordance with those obtained in the original study, where it is shown that Deep-SE achieves lower MAE values than both baseline techniques for all 16 projects (see the \textit{Orig} column in Table~\ref{tbl.replication}) with statistically significant differences in 14 and 15 out of the 16 cases when compared to the Median and Mean estimators, respectively. 

Variation in the results due to use of stochastic techniques, such as deep learning, can be acceptable to some extent \cite{tawosi2021tse}. In fact, the variation we observed in the MAE values achieved for Deep-SE in the original study and this replication can be attributed to such a stochastic nature. However, the difference observed between the results of the baseline techniques reported in the the original study and this replication suggests unjustifiable discrepancies, since the baseline techniques are deterministic and should always yield the same results given the same datasets.
Therefore, we further investigated the origin of such discrepancies by analysing the results and code provided in the original study's replication package and by contacting the authors. We found that the likely cause for these discrepancies is a fault in the computation of the MAE and MdAE of the baseline estimators, and the original study's use of a possibly different approach for random guessing.  
Section~\ref{sec:sp-replication discrepancies} provides a detailed explanation of these causes.

Moreover, while investigating the implementation provided for Deep-SE, we noticed that Deep-SE performs a transformation on the distribution of the SPs in the pre-processing stage (see Section~\ref{sec:sp-replication discrepancies} for more details). This transformation is applied on the SP distribution of each project, before being split into training/validation/test sets, however this pre-processing is not mentioned in the paper. 

To further investigate the effect of such a transformation on the results, we used Deep-SE with the original SPs (i.e., without transforming its distribution) and we report the results in Table~\ref{tbl.replication} under the \textit{!Cut} column.
Our results show that in this case all estimators perform generally worse. Specifically Deep-SE\textit{!Cut} obtains worse MAE values than those it obtains when the transformation is applied in our replication study (\textit{Rep}) in 15 cases (94\%) with the difference being statistically significant in four of them (27\%) and the effect size being small in one and negligible in the remaining three.  While the Mean and Median estimators perform worse in 16 (100\%) and 14 (88\%) cases, respectively with the remaining two cases being equivalent. 
The difference is statistically significant in 12 cases when compared to the Mean estimator with medium effect size in 2 of them, small in 7, and negligible in 2 cases. While, when compared to the Median estimator, the difference is not statistically significant for any of the cases.

We also observe differences between the results shown in Table~\ref{tbl.replication} \textit{!Cut} and the original study (\textit{Orig}): Deep-SE performs worse without applying the SP transformation in all 16 cases considered; whereas, the Mean and Median estimators obtain worse MAE values in 6 (38\%) and 4 (25\%) cases, respectively. 

When we compare the performance of Deep-SE to that of Mean and Median when no transformation is applied i.e., \textit{!Cut}), we observe that Deep-SE outperforms Mean and Median in 15 and 6 cases, respectively. 
Out of the 15 cases, the difference between Deep-SE\textit{!Cut} and Mean is statistically significant in 12 cases, with a medium effect size in 3 and a small one in the remaining 9 cases. As for the Median estimator, out of the 6 cases, only 2 cases show statistically significant difference with small effect size in one and negligible effect size in the other. If we consider the Bonferroni correction, all 12 cases for the Mean estimator are still statistically significant, while the number of statistically significant cases for the Median estimator drops to just one having a small effect size.

We conclude that the transformation induces the techniques to produce a lower absolute error by reducing the range of the SP distribution of the entire dataset (including the test set), therefore suggesting more optimistic results than the ones that could be achieved in practice, where the test set is not available at prediction time. Indeed, when we train Deep-SE while correctly applying the transformation on the training set only (column \textit{CutTrain} in Table~\ref{tbl.replication}), we find that in the majority of the cases (12 out of 16) its performance deteriorates, however the difference is statistically significant in only one of these 12 cases with a negligible effect size. 

In this study, we apply the transformation on the SP distributions for research questions RQ1.1 and RQ5 for replication purposes, while we use the original SP values to answer the remaining RQs (i.e., Deep-SE!Cut is the variant used in the remaining RQs).

\subsubsection{RQ 1.2 - Extension.} 
\label{sec:rq1.2}
To answer RQ 1.2, we benchmark Deep-SE against the Mean and Median estimators using the Tawosi dataset. 

The results reported in Table~\ref{tbl.porru.sp.myds} show that Deep-SE outperforms Mean in 20 out of 26 cases (77\%). However, the difference in absolute error is statistically significant in 16 (62\%) with four cases having a negligible effect size, seven cases having a small effect size, two medium ones, and the remaining three having a large effect size as shown in Table~\ref{tbl.wilcox.porru.sp.myds}. When compared to the Median estimator, Deep-SE obtains lower (better) MAE values in 10 out of the 26 cases investigated (38\%) and only four (15\%) show statistically significant differences with one having a large effect size, one small, and the other two having a negligible effect size. As for the remaining cases, we observe that the Median estimator outperforms Deep-SE in 11 out of 26 projects (42\%) and obtains the same MAE values in five other cases (20\%).
When considering the Bonferroni correction, the number of statistically significant differences do not change for the Mean estimator, while for the Median estimator this number drops to two cases only, one with negligible and one with a large effect size.

Overall, the results obtained for RQs 1.1-RQ1.2 show:

\begin{tcolorbox}
\textbf{Answer to RQ1}: \textit{Deep-SE statistically significantly outperform the baseline estimators in only 42\% of the cases for within-project estimation, and therefore does not pass the sanity check.}
\end{tcolorbox}

\begin{table}
\caption{RQ1.1 and RQ5. Results obtained for the Choet dataset in RQ1.1.: The column ``Rep'' shows the replication results, the column ``Orig" presents the original study results \cite{deep2018}, both obtained by using Deep-SE with the transformed SPs as done in the original study. The column ``CutTrain'' shows the results achieved by applying the transformation only on the training set, while the column ``!Cut'' shows the results of our replication without transforming the SPs.  We also include in this table the results for RQ5 ``Deep-SE!pre-train'', which investigates Deep-SE without pre-training its lower layers (i.e., word embedding and LSTM). Best results in bold.}
\label{tbl.replication}
\centering
\resizebox{\textwidth}{!}{
\begin{tabular}{>{\raggedright\arraybackslash}p{0.42\linewidth} >{\raggedright\arraybackslash}p{0.42\linewidth}  r r r a }
\toprule
Project&Method& Rep  &CutTrain & !Cut & Orig \\
\midrule
MESOS&Deep-SE&1.05&\textbf{1.15}&\textbf{1.12}&\textbf{1.02}\\
&Deep-SE!pre-train&\textbf{1.02}&&&\\
&Mean&1.11&1.22&1.41&1.64\\
&Median&1.11&1.22&1.22&1.73\\
\midrule
USERGRID&Deep-SE&1.06&1.18&1.18&\textbf{1.03}\\
&Deep-SE!pre-train&1.11&&&\\
&Mean&1.09&1.21&1.19&1.48\\
&Median&\textbf{1.03}&\textbf{1.15}&\textbf{1.15}&1.60\\
\midrule
TISTUD&Deep-SE&1.41&1.43&1.42&\textbf{1.36}\\
&Deep-SE!pre-train&1.42&&&\\
&Mean&1.52&1.55&1.91&2.08\\
&Median&\textbf{1.28}&\textbf{1.30}&\textbf{1.30}&1.84\\
\midrule
APSTUD&Deep-SE&3.57&4.09&4.14&\textbf{2.71}\\
&Deep-SE!pre-train&3.15&&&\\
&Mean&\textbf{2.95}&\textbf{3.48}&\textbf{3.59}&3.15\\
&Median&3.08&3.61&3.61&3.71\\
\midrule
TIMOB&Deep-SE&2.10&2.19&2.09&\textbf{1.97}\\
&Deep-SE!pre-train&2.04&&&\\
&Mean&2.53&2.62&3.02&3.05\\
&Median&\textbf{1.94}&\textbf{2.04}&\textbf{2.04}&2.47\\
\midrule
BAM&Deep-SE&0.80&0.80&0.81&\textbf{0.74}\\
&Deep-SE!pre-train&0.77&&&\\
&Mean&1.03&1.03&1.22&1.75\\
&Median&\textbf{0.75}&\textbf{0.75}&\textbf{0.75}&1.32\\
\midrule
CLOV&Deep-SE&2.46&3.75&\textbf{3.39}&\textbf{2.11}\\
&Deep-SE!pre-train&\textbf{2.37}&&&\\
&Mean&2.97&4.26&4.57&3.49\\
&Median&2.42&\textbf{3.71}&3.71&2.84\\
\midrule
JSWSERVER&Deep-SE&\textbf{1.57}&\textbf{1.77}&\textbf{1.70}&\textbf{1.38}\\
&Deep-SE!pre-train&1.58&&&\\
&Mean&1.86&2.07&2.40&2.48\\
&Median&2.10&2.31&2.31&2.93\\
\midrule
DURACLOUD&Deep-SE&\textbf{0.64}&\textbf{0.71}&\textbf{0.82}&\textbf{0.68}\\
&Deep-SE!pre-train&0.68&&&\\
&Mean&0.73&0.82&1.00&1.30\\
&Median&0.76&0.82&0.82&0.73\\
\midrule
DM&Deep-SE&\textbf{3.70}&\textbf{5.88}&\textbf{5.86}&\textbf{3.77}\\
&Deep-SE!pre-train&3.91&&&\\
&Mean&4.89&7.14&8.66&5.29\\
&Median&4.28&6.19&6.19&4.82\\
\midrule
MDL&Deep-SE&6.89&6.89&7.89&\textbf{5.97}\\
&Deep-SE!pre-train&8.05&&&\\
&Mean&10.19&10.19&12.63&10.90\\
&Median&\textbf{6.59}&\textbf{6.59}&\textbf{6.59}&7.18\\
\midrule
MULE&Deep-SE&2.26&2.53&2.59&\textbf{2.18}\\
&Deep-SE!pre-train&2.30&&&\\
&Mean&2.22&2.49&2.60&2.59\\
&Median&\textbf{2.21}&\textbf{2.47}&\textbf{2.47}&2.69\\
\midrule
MULESTUDIO&Deep-SE&\textbf{3.12}&\textbf{3.66}&3.67&\textbf{3.23}\\
&Deep-SE!pre-train&3.24&&&\\
&Mean&3.22&3.70&3.74&3.34\\
&Median&3.18&3.66&\textbf{3.66}&3.30\\
\midrule
XD&Deep-SE&1.66&\textbf{1.63}&\textbf{1.70}&\textbf{1.63}\\
&Deep-SE!pre-train&\textbf{1.58}&&&\\
&Mean&1.88&1.91&2.05&2.27\\
&Median&1.68&1.71&1.71&2.07\\
\midrule
TDQ&Deep-SE&\textbf{2.88}&\textbf{2.90}&3.61&\textbf{2.97}\\
&Deep-SE!pre-train&2.94&&&\\
&Mean&4.08&4.25&4.56&4.81\\
&Median&3.15&3.31&\textbf{3.31}&3.87\\
\midrule
TESB&Deep-SE&\textbf{0.61}&\textbf{0.85}&\textbf{0.90}&\textbf{0.64}\\
&Deep-SE!pre-train&0.63&&&\\
&Mean&0.71&1.00&1.04&1.14\\
&Median&0.70&0.92&0.92&1.16\\
\bottomrule
\end{tabular}
}
\end{table}

\bgroup
\rowcolors{2}{white}{gray!25}
\begin{table}
\caption{RQ1.1 and RQ5. Results of the Wilcoxon test ($\hat{A}_{12}$ effect size in parentheses) comparing Deep-SE vs. baselines (Mean, Median), vs. Deep-SE!pre-train, and vs. Deep-SE!cut, on the Choet dataset.}
\label{tbl.wilcox.replication}
\centering
\resizebox{\columnwidth}{!}{
\begin{tabular}{l r r | r | r}
  \toprule
  \multirow{2}{*}{Project}&\multicolumn{4}{l}{Deep-SE vs.}\\\cmidrule{2-5}
  &Mean&Median&Deep-SE!pre-train&Deep-SE!Cut\\
  \midrule
MESOS & 0.139 (0.52) \_  & 0.182 (0.52) \_  &0.592 (0.49) \_ &0.528 (0.50) \_\\
USERGRID  & 0.440 (0.51) \_  & 0.955 (0.43) \_  &0.318 (0.52) \_  &0.419 (0.51) \_\\
TISTUD  & 0.015 (0.54) n  & 1.000 (0.39) \_  &0.734 (0.49) \_ &0.622 (0.49) \_\\
APSTUD  & 0.979 (0.44) \_  & 0.960 (0.45) \_  &0.953 (0.45) \_ &0.352 (0.51) \_\\
TIMOB & \textless 0.001 (0.60) s  & 0.999 (0.44) \_  &0.893 (0.48) \_ &0.947 (0.47) \_\\
BAM & \textless 0.001 (0.64) s  & 0.886 (0.45) \_  &0.617 (0.49) \_ &0.568 (0.49) \_\\
CLOV  & \textless 0.001 (0.68) s  & 0.063 (0.57) \_  &0.782 (0.46) \_ &0.678 (0.48) \_\\
JSWSERVER & 0.018 (0.60) s  & 0.003 (0.63) s  &0.368 (0.52) \_ &0.637 (0.48) \_\\
DURACLOUD & 0.012 (0.58) n  & 0.936 (0.45) \_  &0.690  (0.48) \_ &0.134 (0.54) \_\\
DM  & \textless 0.001 (0.66) s  & 0.013 (0.53) n  &0.007 (0.53) n &0.019 (0.53) n\\
MDL & \textless 0.001 (0.75) m  & 0.517 (0.50) \_  &\textless0.001  (0.58) n &\textless{0.001} (0.59) n\\
MULE  & 0.572 (0.49) \_ & 0.658 (0.49) \_  &0.532 (0.50) \_ &0.160 (0.53) \_\\
MULESTUDIO  & 0.428 (0.51) \_  & 0.442 (0.50) \_  &0.362 (0.51) \_ &0.370 (0.51) \_\\
XD  & \textless 0.001 (0.58) n  & 0.266 (0.51) \_  &0.991 (0.46) \_ &0.399 (0.50) \_\\
TDQ & \textless 0.001 (0.70) m& 0.009 (0.56) n  &0.219 (0.52) \_ &\textless{0.001} (0.60) s\\
TESB  & \textless 0.001 (0.60) s  & 0.012 (0.57) n  &0.233 (0.52) \_ &0.019 (0.56) n\\
\bottomrule
\end{tabular}
}
\end{table}
\egroup

\bgroup
\rowcolors{2}{white}{gray!25}
\begin{table}
\caption{RQs 1.2, 2.2, and 5. Results of Deep-SE, Deep-SE!pre-train, TF/IDF-SVM, and baseline estimators (Mean, Median) on the Tawosi dataset. Best results in bold.}
\label{tbl.porru.sp.myds}
\centering
\resizebox{\textwidth}{!}{
\begin{tabular}{l | r r r r r}
\toprule
Project& Deep-SE & Deep-SE!pre-train & TF/IDF-SVM & Mean  & Median  \\
\midrule
MESOS&    \textbf{1.34} & 1.43  & \textbf{1.34} & 1.37  & \textbf{1.34} \\
ALOY&   1.51  & 1.71  & \textbf{1.44} & 2.23  & \textbf{1.44} \\
TISTUD&   1.63  & 1.68  & \textbf{1.51} & 2.01  & \textbf{1.51} \\
APSTUD&   4.31  & 4.15  & \textbf{3.99} & 4.00  & \textbf{3.99} \\
CLI&    1.76  & \textbf{1.58} & 2.98  & 2.14  & 1.77  \\
DAEMON&   3.29  & 3.00  & \textbf{2.74} & 2.75  & \textbf{2.74} \\
TIDOC&    \textbf{2.72} & 3.26  & 3.03  & 2.99  & 2.77  \\
TIMOB&    \textbf{2.41} & 2.49  & 2.53  & 2.55  & 2.53  \\
CLOV&   \textbf{3.78} & 3.89  & 4.04  & 5.93  & 4.01  \\
CONFCLOUD&    1.48  & 1.44  & \textbf{1.33} & 1.49  & \textbf{1.33} \\
CONFSERVER&   \textbf{0.91} & 0.96  & 0.96  & 1.35  & 0.96  \\
DNN&    0.72  & 0.72  & 0.79  & 0.80  & \textbf{0.71} \\
DURACLOUD&    0.68  & 0.74  & 0.68  & \textbf{0.67} & 0.68  \\
FAB&    0.86  & 0.75  & 1.10  & 1.19  & \textbf{0.67} \\
STL&    1.18  & 1.20  & \textbf{0.84} & 0.97  & 0.95  \\
DM&   1.61  & 1.65  & \textbf{1.49} & 2.60  & 1.61  \\
COMPASS&    1.63  & 1.66  & \textbf{1.38} & 1.48  & \textbf{1.38} \\
SERVER&   0.89  & 0.87  & 0.93  & 1.56  & \textbf{0.85} \\
EVG&    0.63  & \textbf{0.62} & 0.69  & 0.68  & 0.69  \\
MDL&    \textbf{3.55} & 5.08  & 6.31  & 14.54 & 6.31  \\
MULE&   2.24  & \textbf{2.22} & 3.58  & 2.79  & 2.24  \\
NEXUS&    1.08  & \textbf{1.05} & 1.17  & 1.11  & 1.17  \\
XD&   1.45  & \textbf{1.42} & 2.01  & 1.65  & 1.55  \\
TDP&    0.99  & \textbf{0.98} & 0.99  & 1.17  & 0.99  \\
TDQ&    \textbf{2.47} & 2.92  & 5.05  & 4.20  & 2.88  \\
TESB&   1.15  & 1.09  & \textbf{0.97} & 0.99  & 0.98  \\
\bottomrule
\end{tabular}
}
\end{table}
\egroup

\bgroup
\rowcolors{2}{white}{gray!25}
\begin{table}
\caption{RQs 1.2, 2.2, and 5. Results of the Wilcoxon test ($\hat{A}_{12}$ effect size in parentheses) comparing Deep-SE vs. Deep-SE!pre-train,  Deep-SE vs. TF/IDF-SVM,  Deep-SE vs. baseline estimators (Mean, Median), and TF/IDF-SVM vs. baseline estimators (Mean, Median) on the Tawosi dataset.}
\label{tbl.wilcox.porru.sp.myds}
\centering
\resizebox{1\textwidth}{!}{
\begin{tabular}{l r | r | r r | r r}
\toprule
  \multirow{2}{*}{Project}&\multicolumn{4}{l}{Deep-SE vs.}&\multicolumn{2}{|l}{TF/IDF-SVM vs.}\\\cmidrule{2-7}
  &Deep-SE!pre-train &TF/IDF-SVM & Mean  & Median& Mean  & Median\\
  \midrule
MESOS&0.242 (0.52) \_&0.826 (0.48) \_&0.441 (0.50) \_&0.826 (0.48) \_&0.003 (0.56) n&0.500 (0.50) \_\\
ALOY&0.253 (0.54) \_&0.910 (0.42) \_&\textless 0.001 (0.69) s&0.910 (0.42) \_&\textless 0.001 (0.70) m&0.501 (0.50) \_\\
TISTUD&0.034 (0.53) n&1.000 (0.40) \_&\textless 0.001 (0.62) s&1.000 (0.40) \_&\textless 0.001 (0.63) s&0.500 (0.50) \_\\
APSTUD&0.645 (0.48) \_&0.814 (0.46) \_&0.763 (0.47) \_&0.814 (0.46) \_&0.370 (0.51) \_&0.501 (0.50) \_\\
CLI&0.779 (0.46) \_&\textless 0.001 (0.74) m&0.016 (0.61) s&0.249 (0.54) \_&1.000 (0.31) \_&1.000 (0.26) \_\\
DAEMON&0.520 (0.50) \_&0.659 (0.47) \_&0.618 (0.48) \_&0.659 (0.47) \_&0.629 (0.48) \_&0.502 (0.50) \_\\
TIDOC&0.008 (0.57) n&0.521 (0.50) \_&\textless 0.001 (0.60) s&0.141 (0.53) \_&\textless 0.001 (0.60) s&0.746 (0.48) \_\\
TIMOB&0.272 (0.51) \_&0.303 (0.51) \_&0.050 (0.52) \_&0.303 (0.51) \_&0.032 (0.53) n&0.500 (0.50) \_\\
CLOV&0.070 (0.57) \_&0.817 (0.46) \_&\textless 0.001 (0.85) l&0.026 (0.60) s&\textless 0.001 (0.81) l&0.030 (0.59) n\\
CONFCLOUD&0.427 (0.51) \_&0.822 (0.45) \_&0.497 (0.50) \_&0.822 (0.45) \_&0.029 (0.61) s&0.501 (0.50) \_\\
CONFSERVER&0.359 (0.52) \_&0.869 (0.45) \_&\textless 0.001 (0.65) s&0.869 (0.45) \_&\textless 0.001 (0.65) s&0.501 (0.50) \_\\
DNN&0.642 (0.49) \_&0.848 (0.48) \_&\textless 0.001 (0.57) n&0.370 (0.51) \_&\textless 0.001 (0.67) s&0.614 (0.49) \_\\
DURACLOUD&0.341 (0.52) \_&0.949 (0.42) \_&0.125 (0.56) \_&0.949 (0.42) \_&0.181 (0.55) \_&0.501 (0.50) \_\\
FAB&0.882 (0.44) \_&0.220 (0.54) \_&0.013 (0.61) s&0.995 (0.37) \_&\textless 0.001 (0.68) s&0.997 (0.37) \_\\
STL&0.405 (0.52) \_&0.997 (0.33) \_&0.857 (0.43) \_&0.892 (0.42) \_&\textless 0.001 (0.70) m&0.050 (0.59) \_\\
DM&0.684 (0.49) \_&1.000 (0.39) \_&\textless 0.001 (0.80) l&0.687 (0.49) \_&\textless 0.001 (0.83) l&\textless 0.001 (0.58) n\\
COMPASS&0.331 (0.52) \_&0.898 (0.43) \_&0.665 (0.48) \_&0.898 (0.43) \_&0.229 (0.54) \_&0.501 (0.50) \_\\
SERVER&0.844 (0.46) \_&0.593 (0.49) \_&\textless 0.001 (0.75) m&0.322 (0.52) \_&\textless 0.001 (0.80) l&0.483 (0.50) \_\\
EVG&0.767 (0.49) \_&0.025 (0.53) n&0.019 (0.54) n&0.025 (0.53) n&0.989 (0.46) \_&0.500 (0.50) \_\\
MDL&\textless 0.001 (0.67) s&\textless 0.001 (0.83) l&\textless 0.001 (1.00) l&\textless 0.001 (0.83) l&\textless 0.001 (1.00) l&0.500 (0.50) \_\\
MULE&0.564 (0.50) \_&\textless 0.001 (0.61) s&\textless 0.001 (0.65) s&0.079 (0.52) \_&0.674 (0.49) \_&1.000 (0.39) \_\\
NEXUS&0.853 (0.47) \_&0.847 (0.48) \_&0.539 (0.50) \_&0.847 (0.48) \_&0.573 (0.50) \_&0.500 (0.50) \_\\
XD&0.800 (0.47) \_&0.055 (0.55) \_&0.014 (0.57) n&0.691 (0.48) \_&0.218 (0.52) \_&0.990 (0.43) \_\\
TDP&0.735 (0.47) \_&0.355 (0.52) \_&0.012 (0.59) n&0.355 (0.52) \_&0.008 (0.60) s&0.501 (0.50) \_\\
TDQ&0.003 (0.59) n&\textless 0.001 (0.81) l&\textless 0.001 (0.77) m&0.006 (0.58) n&1.000 (0.35) \_&1.000 (0.21) \_\\
TESB&0.640 (0.49) \_&0.989 (0.42) \_&0.961 (0.44) \_&0.976 (0.43) \_&0.002 (0.59) n&0.406 (0.51) \_\\
\bottomrule
\end{tabular}
}
\end{table}
\egroup

\subsection{RQ2. Deep-SE vs. TF/IDF-SVM} 
\label{sec:Deep-SE vs. TF/IDF-SVM}

\subsubsection{RQ2.1 - Replication} 
The original study evaluated Deep-SE and TF/IDF-SVM on the Porru dataset and found that Deep-SE outperformed TF/IDF-SVM for all the eight projects under investigation.

To answer RQ2.1, we replicate the same experiment with the same data used in the original study. 

Table \ref{tbl.porruds} shows the results we obtained in our replication study using Deep-SE (i.e., \textit{Deep-SE (Rep)}) and \textit{TF/IDF-SVM (Rep)}), together with the results obtained by the original study (\textit{Deep-SE (Orig)} and \textit{TF/IDF-SVM (Orig)}, shaded in grey).

First of all, we observe that in our replication of TF/IDF-SVM we obtain exactly the same results as Choetkiertikul et al. \cite{deep2018} (i.e., the TF/IDF-SVM results are fully reproducible), whereas the results we obtained for Deep-SE are different from those obtained in the original study \cite{deep2018}.
Deep-SE outperforms TF/IDF-SVM in six out of eight projects (75\%) and, among these, the difference is statistically significant in four projects (i.e., MULE, XD, DNN and NEXUS), all with a  small effect size (Table \ref{tbl.wilcox.porru}).
These results contradict those reported in the original study, where Deep-SE was found to be significantly better than TF/IDF-SVM for all eight projects investigated. 

For completeness, although not included in the original work, we also compare the results of Deep-SE and TF/IDF-SVM on the Porru dataset to the Mean and Median baseline estimators.
We can observe that there is a difference in the results of the original study and those we obtain in our replication for all eight projects considered. When compared to the baseline estimators, Deep-SE(Rep) outperforms Mean in seven out of eight (88\%) cases with the differences in four of them being statistically significant and the effect size being negligible in one case, small in one case and large in two other cases. Deep-SE obtains better MAEs than Median in four projects (50\%), out of which two show statistical significance with a small effect size. When we consider the Bonferroni correction, the number of cases where the difference with Mean remains significant drops to three (one with small and two with large effect size). The same observation hold instead for the comparison with the Median estimator under the Bonferroni correction.

\subsubsection{RQ2.2 - Extension}
Table~\ref{tbl.porru.sp.myds} shows the results obtained by \mbox{Deep-SE} and TF/IDF-SVM on the Tawosi dataset.

We observe that Deep-SE performs better than TF/IDF-SVM on 14 out of 26 projects (54\%) and worse on nine projects (34\%). Both perform equally good on the remaining three projects (12\%). However, in the cases where Deep-SE outperforms TF/IDF-SVM, the difference in absolute errors is statistically significant in only five out of 14 cases (19\% of all cases). Among those, the effect size is large in two cases, medium in one, small in one and negligible in one (see Table~\ref{tbl.wilcox.porru.sp.myds}).
For the cases where Deep-SE performs worse than TF/IDF-SVM, the difference in absolute errors is significant in four out of nine cases (15\%), one having a negligible effect size and the remaining three having a small one.

Overall, these results strengthen our confidence in the conclusion of RQ2.1 by confirming a small difference between the results of these two techniques.

\begin{tcolorbox}
\textbf{Answer to RQ2}: \textit{Deep-SE statistically significantly outperforms TF/IDF-SVM in only five out of 26 cases (19\%), whereas it provides statistically significantly worse results in four cases (15\%). Therefore, we cannot conclude that Deep-SE outperforms TF/IDF-SVM for all projects herein.}
\end{tcolorbox}

\begin{table}
\caption{RQ2.1. Results of the Deep-SE and TF/IDF-SVM replication (Rep), original study \cite{deep2018} (Orig), and the baselines on the Porru dataset. Best results (among all methods but Deep-SE (Orig) and TF/IDF-SVM (Orig)) in bold.}
\label{tbl.porruds}
\centering
\resizebox{\textwidth}{!}{
\begin{tabular}{>{\columncolor{white}}l r r r r a a}
  \toprule
Project & \makecell{Deep-SE\\(Rep)} & \makecell{TF/IDF-SVM\\(Rep)} & Mean  & Median  & \makecell{Deep-SE\\(Orig)}  & \makecell{TF/IDF-SVM\\(Orig)}  \\
\midrule
TIMOB & 7.36  & \textbf{1.76} & 20.08 & \textbf{1.76} & 1.44  & 1.76  \\
TISTUD  & 1.36  & \textbf{1.28} & 1.87  & \textbf{1.28} & 1.04  & 1.28  \\
APSTUD  & \textbf{5.52} & 5.69  & 5.59  & 5.69  & 2.67  & 5.69  \\
MESOS & 1.08  & 1.23  & 1.24  & \textbf{0.84} & 0.76  & 1.23  \\
MULE  & 3.27  & 3.37  & 3.22  & \textbf{3.07} & 2.32  & 3.37  \\
XD  & \textbf{1.23} & 1.86  & 1.24  & 1.34  & 1.00  & 1.86  \\
DNN & \textbf{0.69} & 1.08  & 0.72  & 1.08  & 0.47  & 1.08  \\
NEXUS & \textbf{0.30} & 0.39  & 0.72  & 0.39  & 0.21  & 0.39  \\
\bottomrule
\end{tabular}
}
\end{table}

\bgroup
\rowcolors{2}{white}{gray!25}
\begin{table}
\caption{RQ2.1. Results of the Wilcoxon test ($\hat{A}_{12}$ effect size in parentheses) comparing Deep-SE  vs. TF/IDF-SVM,  Deep-SE  vs. baselines (Mean, Median), and TF/IDF-SVM vs. baselines (Mean, Median) on the Porru dataset.}
\label{tbl.wilcox.porru}
\centering
\resizebox{\textwidth}{!}{
\begin{tabular}{l r | r r | r r}
  \toprule
  \multirow{2}{*}{Project}&\multicolumn{3}{l}{Deep-SE vs.}&\multicolumn{2}{|l}{TF/IDF-SVM vs.}\\\cmidrule{2-6}
  &TF/IDF-SVM & Mean  & Median & Mean  & Median\\
  \midrule
TIMOB&0.968 (0.43) \_&\textless 0.001 (0.99) l&0.968 (0.43) \_ & \textless 0.001 (1.00) l & 0.500 (0.50) \_\\
TISTUD&1.000 (0.38) \_&\textless 0.001 (0.67) s&1.000 (0.38) \_& \textless 0.001 (0.64) s & 0.500 (0.50) \_\\
APSTUD&0.366 (0.52) \_&0.413 (0.52) \_&0.366 (0.52) \_& 0.666 (0.48) \_ & 0.501 (0.50) \_\\
MESOS&0.486 (0.50) \_&0.087 (0.57) \_&0.996 (0.37) \_& \textless 0.001 (0.66) s & 0.989 (0.40) \_\\
MULE&0.043 (0.55) s&0.039 (0.56) n&0.080 (0.55) \_& 0.990 (0.43) \_ & 0.997 (0.41) \_\\
XD&0.005 (0.61) s&0.280 (0.53) \_&0.238 (0.53) \_& 0.987 (0.41) \_ & 0.995 (0.39) \_\\
DNN&\textless 0.001 (0.62) s&0.150 (0.53) \_&\textless 0.001 (0.62) s& 1.000 (0.38) \_ & 0.500 (0.50) \_\\
NEXUS&\textless 0.001 (0.63) s&\textless 0.001 (0.91) l&\textless 0.001 (0.64) s& \textless 0.001 (0.76) m & 0.501 (0.50) \_\\
\bottomrule
\end{tabular}
}
\end{table}
\egroup

\subsection{RQ3. Cross-project Estimation} \label{sec:cross-project}

\subsubsection{RQ3.1 - Replication} 
Following the original study, we performed two different types of experiments in order to investigate the effectiveness of Deep-SE for cross-project SP estimation. 

The first experiment is characterised by the fact that Deep-SE is trained by using SP contained only in projects belonging to a same repository of the target project (i.e., within-repository; for example Deep-SE trained on MESOS and tested on USERGRID, where both belong to Apache repository), while in the second experiment we train Deep-SE by using SPs contained only in projects belonging to a different repository (i.e., cross-repository; for example  Deep-SE trained on MESOS from Apache repository and tested on MULE, which belongs to MuleSoft repository). 

We observe that the results we obtained by using Deep-SE both for the within-repository and  cross-repository experiments are different from those recorded in the original study.
Specifically, the MAE values obtained in our within-repository replication are higher (worse) than those reported in the original study in seven out of the eight (88\%) projects (see Table \ref{tbl.within-repo}). 
Whereas, the MAE values we obtain in the cross-repository replication are higher than those reported in the original study only for four out of eight (50\%) projects (see Table \ref{tbl.cross-repo}).
The difference between Rep and Orig might have two reasons. The first  might be the use of the SP transformation in the original study as further explained in Section \ref{sec:sp-replication discrepancies}. The second reason can be the train-validation-test split rates, as two projects are involved in train, validation, and test, and the original study did not mention how they did split the data for this RQ. We kept the same splitting ratio for train and validation as explained in Section \ref{sec:validation}.  

When we compare the results obtained by Deep-SE for cross-project estimation when it is trained with within-repository data (Table \ref{tbl.within-repo}) vs. training it with cross-repository data (Table \ref{tbl.cross-repo}), we can observe that the latter (i.e. training Deep-SE with data from projects belonging to the same repository of the target project) generally provides lower MAE values. Specifically, the MAE values of Deep-SE (Rep) within-repository are lower than those of Deep-SE (Rep) cross-repository in six out of eight cases, whereas in the remaining two they are very close. 
These results corroborate the observation made in the original study, where the within-repository training was found more beneficial than the cross-repository for all cases (i.e., Deep-SE (Orig) within-repository provides always lower MAE than Deep-SE (Orig) cross-repository). This can be explained by the fact that different organisations may apply different policies for SP estimation \cite{Usman2014}. 

While the original study did not benchmark Deep-SE with the Mean and Median baselines for this RQ, we believe these are necessary benchmarks for this scenario as well. The results of this comparison are reported in Tables  \ref{tbl.within-repo} and \ref{tbl.cross-repo}. 
In the within-repository scenario, Deep-SE performs better than the Mean estimator in four out of the eight cases (50\%) studied with  statistical significant difference with a negligible effect size in three and a small one in one case (see last column of Table \ref{tbl.within-repo}). Whereas, Deep-SE outperforms Median in three cases (38\%) with the difference being statistically significant in two of these cases and the effect size being small for one and negligible for the other (see last column of Table \ref{tbl.within-repo}).
In the cross-repository training scenario, Deep-SE statistically significantly outperforms the Mean estimator in six out of eight cases (75\%), with a large effect size in three of them, medium in one, and small in two cases. This is due to the distribution of the SP values in the source and target project. The difference between the MAE values achieved by the Mean estimator and those of Deep-SE are higher for the cases in which the difference in the mean SP values of the source and target projects is larger (see Table \ref{tbl.ourds-porru-stat}).
While, when compared to the Median estimator, our results show that Deep-SE obtains better MAE values in two cases out of eight (25\%) with the difference being statistically significant and the effect size being small in both cases.
If we consider the Bonferroni correction these observations still hold.
Overall, for both scenarios, Deep-SE outperforms the Mean and Median estimators with a statistically significant difference in 10 and 4 out of 16 cases (63\% and 25\%), respectively.

Finally, we comment on the use of Deep-SE for within-project estimation versus its use for cross-project estimation by observing that, overall, Deep-SE is more effective for the former. In fact, the estimation accuracy achieved by Deep-SE on the same set of target projects (i.e., USERGRID, MESOS, APSTUD, TIMOB, TISTUD, MULESTUDIO, MULE when trained based on within-project  -- see Table~\ref{tbl.replication}) is always higher than the accuracy of Deep-SE when trained with cross-project with both within- or cross-repository (see Table~\ref{tbl.cross-project}). This confirms the findings of the original study: Deep-SE is more suitable for within-project estimation than for cross-project estimation.

\subsubsection{RQ3.2 - Extension} 
For this research question, we exploit issues from projects belonging to the same repository as the target project to form a training/validation set. Therefore, we can only use those repositories in the Tawosi dataset that contain more than one project (i.e., Appcelerator, Atlassian, Hyperledger, MongoDB, and Talendforge).
These repositories contain 18 projects, in total.
For each of these 18 projects (as target projects), we combine all the issues from all the other projects belonging to the same repository to form the training/validation set (as the training source).
We then remove the issues that are created after the start date of the target project from the training/validation set. For 13 out of 18 target projects, this leaves us with a source issue set containing less than 200 issues. We remove these target projects from our experiments, as previous work showed training with less than 200 issues may not result in a stable model \cite{porru2016}.
Thus, we investigate five projects for this research question, namely, ALOY with 1,620 issues as the training source, CLI with 3,383 issues, DAEMON with 5,587 issues, TIDOC with 335 issues, and TDP with 1,410 issues.
We split these source issue sets into 75\%-25\% train-validation sets and use them to train Deep-SE. The resulting model is used to estimate all the issues in the target projects.

Table~\ref{tbl.within-repo-rq3.2} reports the results of our experiments. We can observe that, based on the MAE, Deep-SE always provides better results than the Mean estimator but with a statistically significant difference only in three of them. Whereas, Deep-SE performs worse than the Median estimator in all five cases, with the difference being statistically significant in two cases (TIDOC and TDP), and the effect size being negligible. If we consider the Bonferroni correction the same observations hold.

These results corroborate the conclusion drawn in RQ3.1: Deep-SE does not always provide more accurate cross-project SP estimations than baseline estimators. Moreover, our results for RQ3.2 highlight that when Deep-SE is used in a realistic scenario, taking into account the chronological order of the issues, its prediction performance worsens.

\begin{tcolorbox}
\textbf{Answer to RQ3:} \textit{Deep-SE is less effective for cross-project SP estimation with respect to within-project SP estimation.}
\end{tcolorbox}

\begin{table}
\caption{RQ3.1. Comparing Deep-SE cross-project SP estimation replication results (Rep) to the original study results (Orig) \cite{deep2018}, and to the baselines. The results of the Wilcoxon test ($\hat{A}_{12}$ effect size in parentheses) for Deep-SE (Rep) vs. Mean and Median baselines are shown in the last column. Best results (among all approaches but Deep-SE (Orig)) per project  are highlighted in bold.}
\label{tbl.cross-project}
\centering
\begin{subtable}[t]{\textwidth}
    \centering
    \caption{Within-Repository Training}
    \label{tbl.within-repo}
    \resizebox{1\textwidth}{!}{
    \begin{tabular}{>{\raggedright\arraybackslash}p{0.30\linewidth} >{\raggedright\arraybackslash}p{0.30\linewidth} l r r}
    \toprule
    Source&Target&Method&MAE&Deep-SE (Rep) vs.\\
    \midrule
    &&\cellcolor{gray!25}Deep-SE (Orig)&\cellcolor{gray!25}1.07&\\
    MESOS&USERGRID&Deep-SE (Rep)&1.16&\\
    (ME)&(UG)&Mean&1.02&1.000 (0.42) \_\\
    &&Median&\textbf{0.89}&1.000 (0.37) \_\\
    \midrule
    &&\cellcolor{gray!25}Deep-SE (Orig)&\cellcolor{gray!25}1.14&\\
    USERGRID&MESOS&Deep-SE (Rep)&1.51&\\
    (UG)&(ME)&Mean&1.52&0.282 (0.51) \_\\
    &&Median&\textbf{1.50}&0.802 (0.49) \_\\
    \midrule
    &&\cellcolor{gray!25}Deep-SE (Orig)&\cellcolor{gray!25}2.75&\\
    TISTUD&APSTUD&Deep-SE (Rep)&4.37&\\
    (AS)&(AP)&Mean&\textbf{4.27}&0.918 (0.48) \_\\
    &&Median&4.38&0.573 (0.50) \_\\
    \midrule
    &&\cellcolor{gray!25}Deep-SE (Orig)&\cellcolor{gray!25}1.99&\\
    TISTUD&TIMOB&Deep-SE (Rep)&3.38&\\
    (AS)&(TI)&Mean&3.45&\textless 0.001 (0.54) n\\
    &&Median&\textbf{3.17}&1.000 (0.45) \_\\
    \midrule
    &&\cellcolor{gray!25}Deep-SE (Orig)&\cellcolor{gray!25}2.85&\\
    APSTUD&TISTUD&Deep-SE (Rep)&\textbf{2.70}&\\
    (AP)&(AS)&Mean&3.38&\textless 0.001 (0.59) n\\
    &&Median&3.17&\textless 0.001 (0.56) n\\
    \midrule
    &&\cellcolor{gray!25}Deep-SE (Orig)&\cellcolor{gray!25}3.41&\\
    APSTUD&TIMOB&Deep-SE (Rep)&\textbf{3.51}&\\
    (AP)&(TI)&Mean&4.36&\textless 0.001 (0.64) s\\
    &&Median&4.19&\textless 0.001 (0.62) s\\
    \midrule
    &&\cellcolor{gray!25}Deep-SE (Orig)&\cellcolor{gray!25}3.14&\\
    MULE&MULESTUDIO&Deep-SE (Rep)&3.64&\\
    (MU)&(MS)&Mean&3.34&0.775 (0.49) \_\\
    &&Median&\textbf{3.26}&0.997 (0.46) \_\\
    \midrule
    &&\cellcolor{gray!25}Deep-SE (Orig)&\cellcolor{gray!25}2.31&\\
    MULESTUDIO&MULE&Deep-SE (Rep)&2.77&\\
    (MS)&(MU)&Mean&3.05&0.004 (0.54) n\\
    &&Median&\textbf{2.60}&0.997 (0.46) \_\\
    \bottomrule
    \end{tabular}
    }
\end{subtable}
\newline
\newline
\begin{subtable}[t]{\textwidth}
    \centering
    \caption{Cross-Repository Training}
    \label{tbl.cross-repo}
    \resizebox{\textwidth}{!}{
    \begin{tabular}{>{\raggedright\arraybackslash}p{0.30\linewidth} >{\raggedright\arraybackslash}p{0.30\linewidth} l r r r r}
    \toprule
    Source&Target&Method&MAE&Deep-SE (Rep) vs.\\
    \midrule
    &&\cellcolor{gray!25}Deep-SE (Orig)&\cellcolor{gray!25}1.57&\\
    TISTUD&USERGRID&Deep-SE (Rep)&3.47&\\
    (AS)&(UG)&Mean&3.08&1.000 (0.42) \_\\
    &&Median&\textbf{2.30}&1.000 (0.27) \_\\
    \midrule
    &&\cellcolor{gray!25}Deep-SE (Orig)&\cellcolor{gray!25}2.08&\\
    TISTUD&MESOS&Deep-SE (Rep)&3.18&\\
    (AS)&(ME)&Mean&3.28&0.011 (0.52) n\\
    &&Median&\textbf{2.58}&1.000 (0.39) \_\\
    \midrule
    &&\cellcolor{gray!25}Deep-SE (Orig)&\cellcolor{gray!25}5.37&\\
    MDL&APSTUD&Deep-SE (Rep)&5.03&\\
    (MD)&(AP)&Mean&9.84&\textless 0.001 (0.81) l\\
    &&Median&\textbf{3.97}&1.000 (0.43) \_\\
    \midrule
    &&\cellcolor{gray!25}Deep-SE (Orig)&\cellcolor{gray!25}6.36&\\
    MDL&TIMOB&Deep-SE&\textbf{3.34}&\\
    (MD)&(TI)&Mean&11.19&\textless 0.001 (0.92) l\\
    &&Median&4.19&\textless 0.001 (0.63) s\\
    \midrule
    &&\cellcolor{gray!25}Deep-SE (Orig)&\cellcolor{gray!25}5.55&\\
    MDL&TISTUD&Deep-SE (Rep)&\textbf{2.64}&\\
    (MD)&(AS)&Mean&11.45&\textless 0.001 (0.97) l\\
    &&Median&3.17&\textless 0.001 (0.58) n\\
    \midrule
    &&\cellcolor{gray!25}Deep-SE (Orig)&\cellcolor{gray!25}2.67&\\
    DM&TIMOB&Deep-SE (Rep)&3.81&\\
    &(TI)&Mean&5.61&\textless 0.001 (0.72) m\\
    &&Median&\textbf{3.46}&1.000 (0.45) \_\\
    \midrule
    &&\cellcolor{gray!25}Deep-SE (Orig)&\cellcolor{gray!25}4.24&\\
    USERGRID&MULESTUDIO&Deep-SE (Rep)&3.95&\\
    (UG)&(MS)&Mean&4.04&0.008 (0.54) n\\
    &&Median&\textbf{3.91}&0.917 (0.48) \_\\
    \midrule
    &&\cellcolor{gray!25}Deep-SE (Orig)&\cellcolor{gray!25}2.70&\\
    MESOS&MULE&Deep-SE (Rep)&3.20&\\
    (ME)&(MU)&Mean&\textbf{2.89}&0.999 (0.46) \_\\
    &&Median&2.92&1.000 (0.45) \_\\
    \bottomrule
    \end{tabular}
    }
\end{subtable}
\end{table}

\begin{table}
\caption{RQ3.2. Comparing the cross-project prediction accuracy (in terms of MAE) of Deep-SE and the baselines Mean and Median. The last column shows the result of the Wilcoxon statistical test ($\hat{A}_{12}$ Effect size in parentheses) for Deep-SE. Best results in bold.}
\label{tbl.within-repo-rq3.2}
\centering
\resizebox{\textwidth}{!}{
    \begin{tabular}{>{\raggedright\arraybackslash}p{0.4\linewidth} l l r r}
    \toprule
    Source&Target&Method&MAE&Deep-SE vs.\\
    \midrule
    APSTUD, &ALOY&Deep-SE &2.15&\\
    TIDOC, TIMOB, &&Mean&2.83&\textless 0.001 (0.67) s\\
    TISTUD &&Median  &\textbf{2.10}&0.770 (0.48) \_\\
    \midrule
    ALOY, APSTUD,&CLI &Deep-SE &2.72&\\
    TIDOC, TIMOB,&&Mean&2.80&0.089 (0.53) \_\\
    TISTUD&&Median  &\textbf{2.38}&0.940 (0.46) \_\\
    \midrule
    ALOY, CLI,&DAEMON  &Deep-SE &2.95&\\
    APSTUD, TIDOC,&&Mean&3.04&0.166 (0.53) \_\\
    TIMOB, TISTUD&&Median  &\textbf{2.89}&0.526 (0.50) \_\\
    \midrule
    APSTUD, TIMOB,&TIDOC&Deep-SE &2.53&\\
    TISTUD&&Mean&2.91&\textless 0.001 (0.64) s\\
    &&Median  &\textbf{2.45}&0.006 (0.53) n\\
    \midrule
    TDQ, TESB&TDP &Deep-SE &1.60&\\
    &&Mean&2.46&\textless 0.001 (0.75) m\\
    &&Median  &\textbf{1.53}&0.021 (0.54) n\\
    \bottomrule
    \end{tabular}
    }
\end{table}

\subsection{RQ4. Augmented Training Set}

To augment the training sets used in RQ1.2, we exploit issues from projects belonging to the same repository as the target project in the Tawosi dataset. 
Similarly to RQ3.2, this limits us to 18 projects in the Appcelerator, Atlassian, Hyperledger, MongoDB, and Talendforge repositories.

Applying the augmentation resulted in an increase in the size of the training set for all 18 target projects, ranging between 63.2 times for TIDOC and 1.4 times for SERVER (10.5 times on average over 18 projects).
This experiment mimics a real-world scenario in which a company uses issues from all its projects (past and present) to train Deep-SE and use it for story point estimation of new issues.

Note that, this is different from RQ3.2, where Deep-SE did not use any issues from the target project for training, while herein, we keep the original 60\%-20\% train-validation sets from the target project and only augment the training set with available issues from other projects\footnote{Note that we only include those issues from other projects that are resolved before the creation of the first issue in the test set.}. The purpose of this is to examine whether Deep-SE's estimation performance will increase, if more training data were available.

Table~\ref{tbl.aug.sp.myds} shows the results we obtained after augmenting the training set for 18 projects (indicated by ``AUG'' as the column header). 
This table also shows the results of Deep-SE estimating SP for the same projects by only using within-project issues (under WP column).
We also report the results of the baseline estimators, to verify whether augmenting the training set helps Deep-SE outperform the baselines on those projects in which Mean and Median had previously performed better in within-project estimation.
Results show that Deep-SE performs better on only five out of 18 projects (28\%), when the training set is augmented. On the other hand, its performance deteriorates on 12 projects (67\%) and remains the same on a single project (CONFCLOUD).

We looked for common features among the projects that are benefited from the augmentation or those which did not, but we could not find any emerging pattern. Specifically, we looked into their application domain, project size, amount of augmented data with respect to the original size of the project, and the respective repository.  

The analysis of the Wilcoxon test on Deep-SE for within-project vs augmentation showed that the difference in the distribution of the errors produced by the two variants is statistically significant in five projects (ALOY, TIDOC, TISTUD, EVG, TDQ) with a medium effect size in ALOY and negligible with the others.

The performance of Deep-SE with respect to Mean and Median remains almost unchanged when the augmented data is used. In fact, on the one hand Deep-SE (AUG) outperforms the baselines on five projects (i.e., ALOY, COMPASS, SERVER, TDP, and TESB), while Deep-SE (WP) was outperformed by the baselines on these five projects (see RQ2.2). However, on the other hand, the performance of Deep-SE (AUG) becomes worse than that of the baseline techniques on four projects (i.e., TISTUD, CONFSERVER, EVG, TDQ), while Deep-SE (WP) performs better than the baselines on these same four projects (see RQ2.2).

\begin{tcolorbox}
\textbf{Answer to RQ4}: \textit{Augmenting the training set with issues from within-company projects had no steady positive or negative effect on Deep-SE's accuracy.}
\end{tcolorbox}

\begin{table}
\caption{RQ4. Results achieved by Deep-SE on the Tawosi dataset when the training set is augmented by using older issues from the repository that the project belongs to (AUG), compared to Deep-SE's within-project results from RQ1.2 (WP), and to baseline estimators. Best results in bold.}
\label{tbl.aug.sp.myds}
\centering
\resizebox{\textwidth}{!}{
\begin{tabular}{l l r r l l r r }
  \toprule
  \multirow{2}{*}{Project}&\multirow{2}{*}{Method}&\multicolumn{2}{c}{MAE}&\multirow{2}{*}{Project}&\multirow{2}{*}{Method}&\multicolumn{2}{c}{MAE}\\
  \cmidrule{3-4}\cmidrule{7-8}
  & & AUG  & WP & & & AUG  & WP  \\
\midrule
ALOY& Deep-SE & \textbf{2.59} & 1.51& CONFSERVER& Deep-SE & 1.04& \textbf{0.91} \\
& Mean& 3.13& 2.23& & Mean& 1.95& 1.35\\
& Median& 2.80& \textbf{1.44} & & Median& \textbf{0.96} & 0.96\\
\midrule
TISTUD& Deep-SE & 1.73& 1.63& FAB & Deep-SE & 0.87& 0.86\\
& Mean& 1.91& 2.01& & Mean& 1.00& 1.19\\
& Median& \textbf{1.51} & \textbf{1.51} & & Median& \textbf{0.67} & \textbf{0.67} \\
\midrule
APSTUD& Deep-SE & 4.49& 4.31& STL & Deep-SE & 1.10& 1.18\\
& Mean& \textbf{4.02} & 4.00& & Mean& 1.24& 0.97\\
& Median& 4.02& \textbf{3.99} & & Median& \textbf{0.95} & \textbf{0.95} \\
\midrule
CLI & Deep-SE & \textbf{2.04} & \textbf{1.76} & COMPASS & Deep-SE & \textbf{1.43} & 1.63\\
& Mean& 3.19& 2.14& & Mean& 1.89& 1.48\\
& Median& 2.98& 1.77& & Median& 1.81& \textbf{1.38} \\
\midrule
DAEMON& Deep-SE & 3.10& 3.29& SERVER& Deep-SE & \textbf{0.83} & 0.89\\
& Mean& 2.76& 2.75& & Mean& 0.85& 1.56\\
& Median& \textbf{2.74} & \textbf{2.74} & & Median& 0.85& \textbf{0.85} \\
\midrule
TIDOC & Deep-SE & \textbf{3.35} & \textbf{2.72} & EVG & Deep-SE & 0.68& \textbf{0.63} \\
& Mean& 3.47& 2.99& & Mean& 0.66& 0.68\\
& Median& 3.42& 2.77& & Median& \textbf{0.65} & 0.69\\
\midrule
TIMOB & Deep-SE & \textbf{2.46} & \textbf{2.41} & TDP & Deep-SE & \textbf{1.07} & \textbf{0.99} \\
& Mean& 2.62& 2.55& & Mean& 2.27& 1.17\\
& Median& 2.53& 2.53& & Median& 1.47& \textbf{0.99} \\
\midrule
CLOV& Deep-SE & \textbf{3.71} & \textbf{3.78} & TDQ & Deep-SE & 2.83& \textbf{2.47} \\
& Mean& 5.35& 5.93& & Mean& 2.82& 4.20\\
& Median& 4.01& 4.01& & Median& \textbf{2.22} & 2.88\\
\midrule
CONFCLOUD & Deep-SE & 1.48& 1.48& TESB& Deep-SE & \textbf{1.19} & 1.15\\
& Mean& 2.32& 1.49& & Mean& 2.52& 0.99\\
& Median& \textbf{1.33} & \textbf{1.33} & & Median& 1.29& \textbf{0.98} \\
\bottomrule
\end{tabular}
}
\end{table}

\subsection{RQ5. Pre-Training Effectiveness}
Tables~\ref{tbl.replication} and \ref{tbl.porru.sp.myds} show the performance of \mbox{Deep-SE} with and without pre-training (Deep-SE vs Deep-SE!pre-train) on the Choet and the Tawosi datasets, respectively. 
We can observe that using pre-training does not always improve Deep-SE's estimation accuracy.
Specifically, Deep-SE with random initialization (i.e., Deep-SE!pre-train) obtained better MAE values, although with a small difference, for six out of 16 (38\%) projects from the Choet dataset (TIMOB, APSTUD, BAM, CLOV, MESOS, XD).

The results of the Wilcoxon test (Table~\ref{tbl.wilcox.replication}) show that the difference in the estimation accuracy of Deep-SE with and without pre-training is statistically significant (in favour of pre-training) for two of the projects (DM and MDL) with a negligible effect size.\footnote{Note that on the other hand, the difference in the estimation accuracy is statistically significant in favour of random initialization (without pre-training) for two project (APSTUD and XD), however, with a negligible effect size.}
On the Tawosi dataset (Table~\ref{tbl.porru.sp.myds}), Deep-SE without pre-training performs similarly or better than Deep-SE in 13 out of 26 projects (50\%) among which the difference in errors produced by the two variants is statistically significant on four projects only (TISTUD, TIDOC, MDL, TDQ) with a negligible/small effect size (Table~\ref{tbl.wilcox.porru.sp.myds}).

Having seen the trivial effect of pre-training in improving the accuracy of Deep-SE for SP estimation, we checked whether it at least helps Deep-SE converge faster to the best solution.
To this end, we compare the running time and the number of epochs (Table~\ref{tbl.deepse.pretrain.converge}) required by the two variants to converge to the best solution on the validation set before the early stopping criterion is met (i.e., ten consecutive epochs with no improvements in the validation loss function).

We can observe that the running time of Deep-SE with random initialization (i.e., Deep-SE!pre-train) is slightly smaller than its pre-trained variant in 33 out of the 42 cases (79\%) studied.
When checking for statistical difference between the distribution of the running time of the two variants, the results of the Wilcoxon test show $p-value=0.956$ for the Choet dataset and $p-value=0.791$ for the Tawosi dataset, indicating that we cannot accept the alternative hypothesis that these results are statically significant different.
Similarly, the comparison between the number of epochs does not reveal any significant difference ($p-value=0.830$), although on average, Deep-SE without pre-training took slightly fewer epochs to converge (see Table~\ref{tbl.deepse.pretrain.converge}).

Overall, the results show that using the pre-trained embedding weights and LSTM layer do not enhance Deep-SE's accuracy or convergence speed; thus, it can be skipped to save the high cost required by such a pre-training procedure.

\bgroup
\rowcolors{3}{white}{gray!25}
\begin{table}
\caption{RQ5. Comparing running time and number of epochs that each of the methods (i.e., Deep-SE when used with initialization of weights through pre-training (Deep-SE) and Deep-SE with random initialization (Deep-SE!pre-train)) needs to converge, on the (a) Choet and (b) Tawosi datasets. Best results in bold.}
\label{tbl.deepse.pretrain.converge}
\centering
\begin{subtable}[t]{\textwidth}
    \centering
    \caption{Choet dataset}
    \label{tbl.deepse.choet.pretrain.converge}
    \resizebox{0.9\textwidth}{!}{
    \begin{tabular}{l | r r | r r}
    \toprule
    \multirow{2}{*}{Project}&\multicolumn{2}{l}{Running Time (Seconds)}&\multicolumn{2}{l}{Epoch}\\\cmidrule{2-5}
    &Deep-SE &Deep-SE!pre-train&Deep-SE &Deep-SE!pre-train\\
    \midrule
    MESOS  & 757  & \textbf{702}  & 115 & \textbf{109} \\
    USERGRID & \textbf{240}  & 260  & \textbf{114} & 129 \\
    TISTUD  & \textbf{1,212} & 1,268 & \textbf{109} & 111 \\
    APSTUD  & 464  & \textbf{363}  & 133 & \textbf{110} \\
    TIMOB & 950  & \textbf{905}  & 109 & \textbf{106} \\
    BAM & 243  & \textbf{243}  & \textbf{107} & 108 \\
    CLOV & 210  & \textbf{204}  & 123 & \textbf{115} \\
    JSWSERVER  & \textbf{186}  & 279  & \textbf{113} & 144 \\
    DURACLOUD  & 329  & \textbf{320}  & 120 & \textbf{117} \\
    DM  & 2,128 & \textbf{1,973} & 119 & \textbf{110} \\
    MDL & 566  & \textbf{522}  & 121 & \textbf{116} \\
    MULE & 429  & \textbf{415}  & 117 & \textbf{111} \\
    MULESTUDIO & 320  & \textbf{310}  & 110 & \textbf{107} \\
    XD  & \textbf{1,433} & 1,475 & \textbf{108} & 109 \\
    TDQ  & 673  & \textbf{629}  & \textbf{111} & 112 \\
    TESB & \textbf{435}  & 452  & 124 & \textbf{123} \\
    \midrule
    \rowcolor{white}
    Total &10,583  & \textbf{10,326}&1,853 & \textbf{1,837}\\
    \midrule
    Mean  & - & - & 115.81 &  \textbf{114.81}\\
    \bottomrule
\end{tabular}
}
\end{subtable}
\newline
\newline
\begin{subtable}[t]{\textwidth}
    \centering
    \caption{Tawosi dataset}
    \label{tbl.deepse.myds.pretrain.converge}
    \resizebox{0.9\textwidth}{!}{
    \begin{tabular}{l | r r | r r}
    \toprule
    \multirow{2}{*}{Project}&\multicolumn{2}{l}{Running Time (Seconds)}&\multicolumn{2}{l}{Epoch}\\\cmidrule{2-5}
    &Deep-SE &Deep-SE!pre-train&Deep-SE &Deep-SE!pre-train\\
    \midrule
    MESOS&844&\textbf{750}&\textbf{111}&116\\
    ALOY&\textbf{144}&159&\textbf{111}&119\\
    TISTUD&1,641&\textbf{1,376}&\textbf{106}&\textbf{106}\\
    APSTUD&317&\textbf{251}&122&\textbf{111}\\
    CLI&226&\textbf{164}&130&\textbf{108}\\
    DAEMON&198&\textbf{149}&131&\textbf{128}\\
    TIDOC&601&\textbf{486}&\textbf{108}&115\\
    TIMOB&2,178&\textbf{1,793}&\textbf{110}&112\\
    CLOV&239&\textbf{217}&125&\textbf{111}\\
    CONFCLOUD&179&\textbf{150}&117&\textbf{108}\\
    CONFSERVER&303&\textbf{261}&112&\textbf{109}\\
    DNN&1,193&\textbf{892}&117&\textbf{106}\\
    DURACLOUD&272&\textbf{217}&\textbf{137}&147\\
    FAB&214&\textbf{172}&118&\textbf{107}\\
    STL&161&\textbf{133}&116&\textbf{115}\\
    DM&2,942&\textbf{2,711}&\textbf{110}&\textbf{110}\\
    COMPASS&259&\textbf{166}&171&\textbf{117}\\
    SERVER&\textbf{276}&307&\textbf{107}&129\\
    EVG&2,421&\textbf{1,967}&\textbf{173}&175\\
    MDL&822&\textbf{719}&\textbf{112}&126\\
    MULE&1,586&\textbf{1,495}&\textbf{111}&114\\
    NEXUS&\textbf{604}&626&\textbf{106}&\textbf{106}\\
    XD&466&\textbf{380}&110&\textbf{109}\\
    TDP&312&\textbf{239}&\textbf{110}&\textbf{110}\\
    TDQ&\textbf{409}&434&\textbf{115}&117\\
    TESB&466&\textbf{350}&119&\textbf{110}\\
    \midrule
    \rowcolor{white}
    Total& 19,273 & \textbf{16,564} & 3,115 & \textbf{3,041} \\
    \midrule
    Mean& - & - &119.81&\textbf{116.96}\\
    \bottomrule
    \end{tabular}
    }
\end{subtable}
\end{table}
\egroup

\begin{tcolorbox}
\textbf{Answer to RQ5}: \textit{Pre-training the lower layer of Deep-SE with user stories from other projects has a negligible effect in improving its SP estimation accuracy or convergence speed; thus, it can be skipped.}
\end{tcolorbox}

\section{Discussion} 
\label{sec:sp-replication discrepancies}

Our replication of the work by Choetkiertikul et al. \cite{deep2018} provided different results from the original study in some cases.
In RQ1.1, we found that the results we achieved for Deep-SE and the baselines are different from that of the original study. Moreover, in RQ2.1, although the results obtained in our study for TF/IDF-SVM matched those reported in the original study, we found differences in the results of Deep-SE between the two studies. Finally, in RQ3.1, we observed differences in the results of Deep-SE between our replication and the original study, although they support the same conclusion as the one made in the original study.

In order to investigate the causes of the discrepancies we: (i) peer-reviewed the modifications and fixes we performed on the original source code and results to make sure it is correct (i.e., the second author reviewed the code written by the first author, and double-checked the results), (ii) checked the source code and results included in the replication package provided by the authors \cite{deepseImpl}; (iii) contacted the authors of the original study.
We discuss below the possible reasons for observing different results.

First of all, we notice a discrepancy in the results reported for the baseline estimators, namely Mean and Median. These na\"ive baselines are deterministic, and therefore we do expect them to achieve the same results when applied to the same data. However, this was not the case, and further investigation
suggests that there may have been a misuse of these estimators in the original study. 
Indeed, the results of the original study are reproducible only by adding one (1.0) to the Mean and Median story point estimates. Since this addition modifies the original estimates of the baselines, it leads to an incorrect absolute error computation on the test set. 
This modification of the original estimates is not described or justified in the original paper, neither in the more general effort estimation literature, and as such we conclude that it should not be used. 
Therefore, for all the RQs investigated in this paper, we report the correct results for the Mean and Median estimators based on the definition given in the literature as described in Section~\ref{sec:evaluation}.

By investigating the source code made publicly available by the authors \cite{deepseImpl}, we also found a discrepancy between the description of Deep-SE provided in the paper and its implementation as included in their replication package.
In fact the Python implementation for Deep-SE applies a transformation on the SP distribution before using it for training, validation, and testing.
Through this transformation, all the original story points falling above a certain threshold (i.e., 90$^{th}$ percentile of the distribution) are replaced with a smaller cap value (specifically, with the value fallen in the 90$^{th}$ percentile). With this transformation in place, for example, in the DM project, 319 issues with SP values between 25 and 100, are all set to 21 in the training set.  The same is done for 68 issues from the test set.

The following caveats arise using such a transformation:
(i) it is not mentioned in the original paper \cite{deep2018}, which can render future replications and/or adoption of the method non-consistent or unreliable;
(ii) our results (RQ1) show that without applying the pre-processing step, an estimation technique tends to perform generally worse.
(iii) the current source code applies the pre-processing step to the entire dataset (i.e., before splitting it into training, validation, and test sets). However, while applying such a transformation on the training set might be acceptable, it should not be applied on the test set since it can undermine the validation process by inflating its accuracy, given that in real world settings, the test set is unknown and such a transformation cannot be applied at prediction time. The corresponding author of the original study explained to us that they performed this pre-processing to investigate the effect of eliminating outliers from the training set, but it was not their intention to apply it to the test set.

Therefore, we conclude that using the code as provided in the replication package might lead users, unaware of the application of such a transformation in the code, to unjustly/unfairly compare Deep-SE with other approaches which do not include this pre-processing step. This may be the case of a recent work by Abadeer and Sabetzadeh \cite{abadeer2021machine} that used Deep-SE with a dataset of industrial projects. Using the Deep-SE's implementation as provided in the replication package, will automatically apply the transformation to the entire dataset before training, validation, and testing, and it will replace all those SPs with a value of 13 (which are 4\% of the issues) with the value 8. Therefore, Deep-SE may produce a lower (better) MAE than the baselines in case they are used on non-transformed data.
Based on the above observations, for replication purposes we reported the result of Deep-SE both with and without transformed SPs for RQ1.1, and with transformation for RQ5. Whereas, for RQ1.2, RQ2, RQ3, and RQ4, we report the results of Deep-SE without transformation.

\begin{table*}
\caption{Semantically related user stories with different SP values for Spring XD. Related concepts highlighted in bold.}
\label{tbl.samplestories}
\centering
\resizebox{0.85\textwidth}{!}{
\begin{tabular}{l l p{0.3\linewidth} p{0.6\linewidth} r}
\toprule
Issue Key&Type&Title&Description&Story Point\\
\midrule
XD-2347&Technical Task&\textbf{Document} \textbf{Kafka message bus}&As a user, I'd like to refer to \textbf{document}ation in wiki so that I can setup and \textbf{configure} \textbf{Kafka} as a \textbf{message bus} as recommended.&2\\
\midrule
XD-2361&Story&Pre-allocate \textbf{partition}s for \textbf{Kafka message bus}&As a user, I want Spring XD’s \textbf{message bus} to be able to pre-allocate \textbf{partition}s between nodes when a stream is deployed, so that rebalancing doesn’t happen when a container crashes and/or it’s redeployed.&8\\
\midrule
XD-3164&Story&\textbf{Kafka bus} defaults \textbf{configurable} at producer/consumer level&As a developer, I want to be able to override \textbf{Kafka bus} defaults for module consumers and producers, so that I can finely tune perfo CT grtrtt  rmance and behaviour. Such \textbf{properties} should include -autoCommitEnabled, queueSize, \textbf{maxWait}, fetchSize for consumers- batchSize, batchTimeout for producers.&3\\
\midrule
XD-3516&Story&\textbf{Document} \textbf{partition}ing through deployment \textbf{properties}&As an s-c-d user, I'd like to have \textbf{document}ation on deployment manifest, so I could refer to the relevant bits on {{\textbf{partition}s}}. I'd like to understand how stream with e... &2\\
\midrule
XD-3740&Bug&\textbf{Kafka message bus} \textbf{maxWait} \textbf{property} is not set up&The \textbf{maxWait property} from server.yml in the \textbf{message bus} section for \textbf{kafka} is not propagated through the code, it is ignored.&1\\
\bottomrule
\end{tabular}
}
\end{table*}

\section{Conclusions}  \label{sec:conclusion}
Considering the variations observed between the replication and the original study \cite{deep2018}, we cannot confirm all conclusions made in the original study for RQs1-2. We found that Deep-SE do not outperform the Median baseline for all projects (RQ1) and is not always statistically significantly better than TF/IDF-SVM (RQ2).
Moreover, the statistical test performed on the distribution of the absolute errors obtained by Median estimator vs. Deep-SE and TF/IDF-SVM, respectively, showed that the null hypothesis cannot be rejected (i.e. the MAEs are different) for over half of the projects investigated (RQ2.1).
A broader experiment with the Tawosi dataset (RQ2.2), which contains a larger number of projects and issues, showed that the results of the these methods are significantly different in only one-fifth of the cases.

Furthermore, we observed that Deep-SE is not so effective for cross-project estimation (RQ3), thus confirming the conclusion made in the original study. Using other additional projects for training Deep-SE or augmenting the training set with issues from  other projects developed within a same repository does not improve its prediction performance statistically significantly (RQ4). We also showed that the pre-training of the lower layers of Deep-SE, which demands a large number of issues and takes a long time to execute, does not have a significant effect on its accuracy nor convergence (RQ5).

\begin{figure*}
\centering
    \begin{subfigure}{0.50\textwidth}
      \includegraphics[width=1\textwidth]{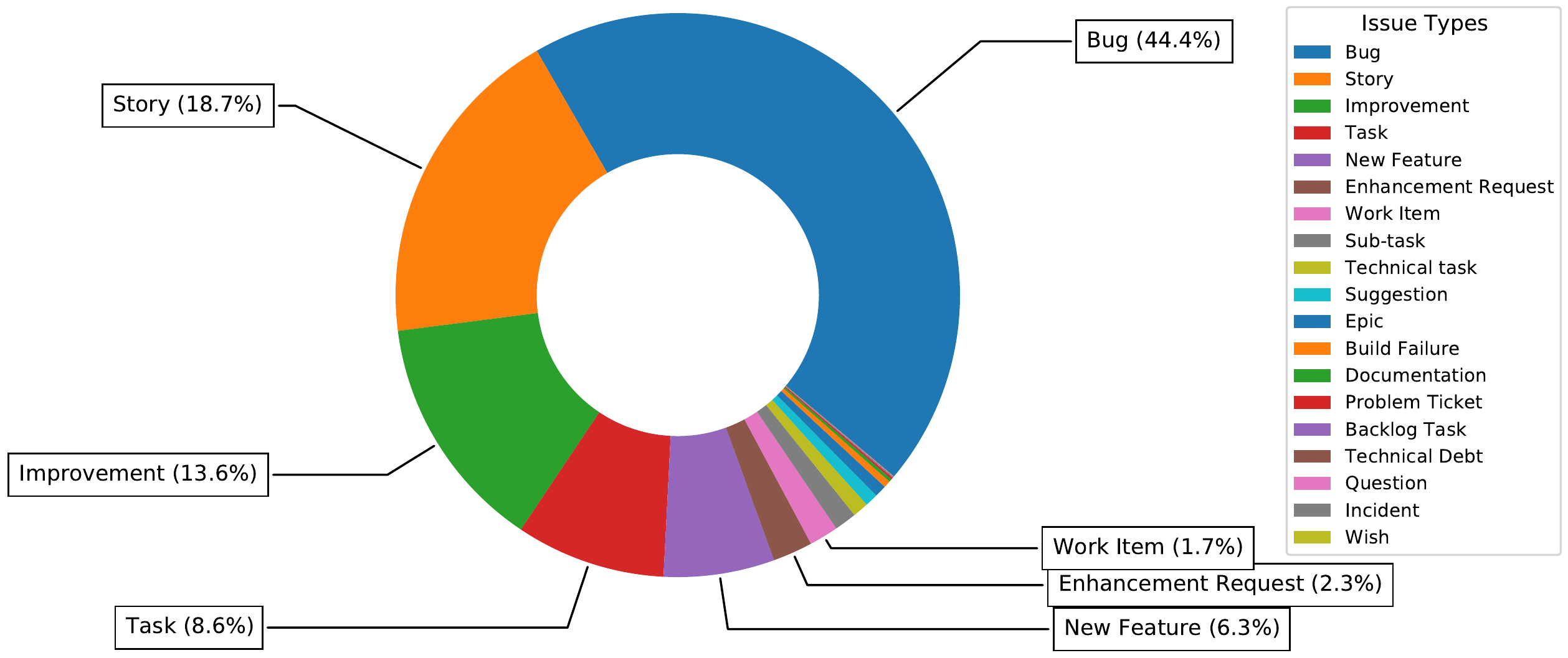}
      \caption{}
      \label{fig_issue_type_pie}
    \end{subfigure}
    \begin{subfigure}{0.60\textwidth}
      \includegraphics[width=1\textwidth]{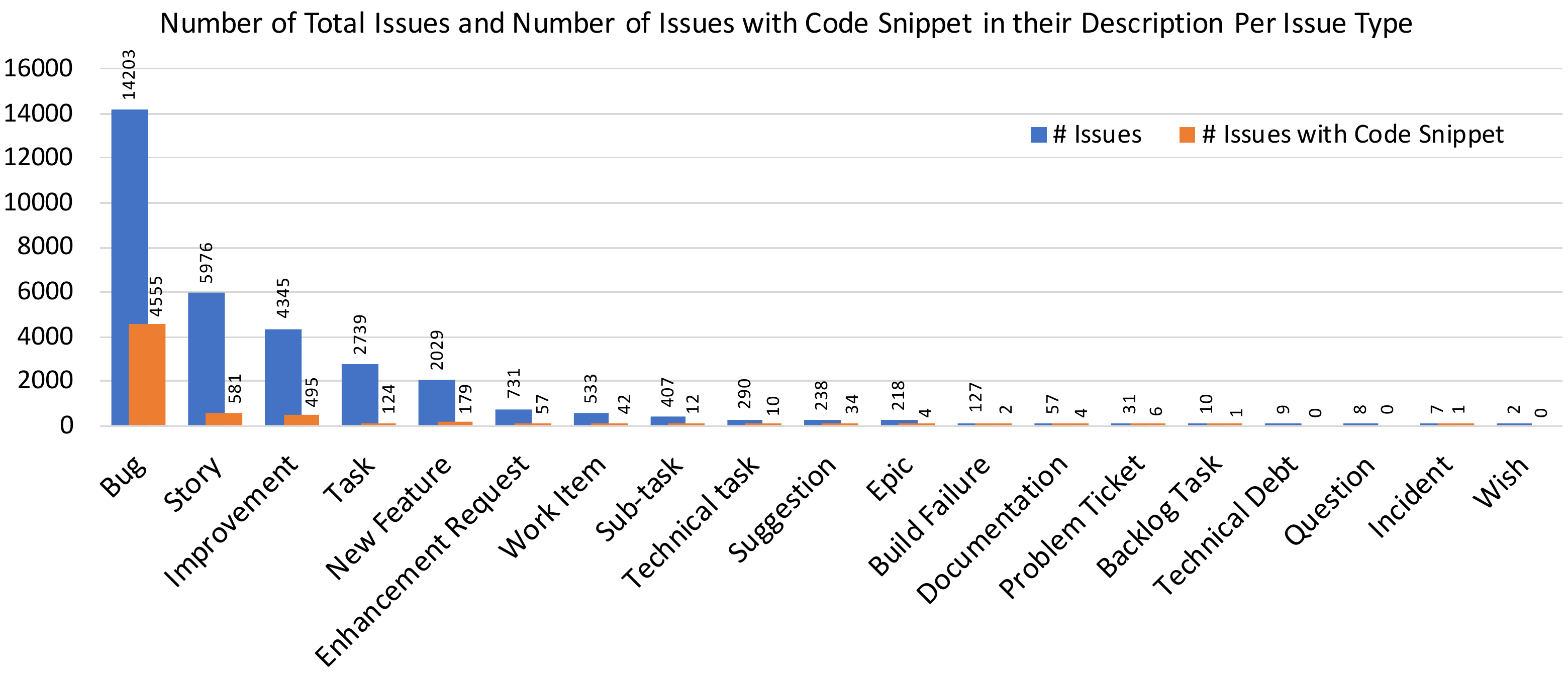}
      \caption{}
      \label{fig_issue_code_bar}
  \end{subfigure}
  \caption{Proportion of Issue Types in the Tawosi Dataset (a), and number of total issues and number of issues with code snippet in their description, grouped by issue type, in the Tawosi dataset (b). Issue types are ordered by their frequency in descending order from left to right. }
\end{figure*}

Overall, our study reveals that current approaches to Agile Software Effort Estimation, 
which in essence try to find semantic similarities between user stories to estimate effort, very often fail to provide statistically significantly better estimations than a na\"ive baseline technique like the Median estimator. This suggests that semantic similarity between user stories might not be sufficient, or even effective, for issue-level effort estimation. For instance, it is possible that two user stories that discuss the same concept (thus are semantically related) demand different amounts of effort to be resolved (e.g., one asks for a new feature to be implemented and the other requires a small change in an already implemented feature).
Table~\ref{tbl.samplestories} shows an example of five semantically related user stories from the Spring XD project. As we can see, these semantically similar user stories are estimated to have different story points. It means that, the distribution of the story points in the groups of semantically related user stories is not necessarily different from the distribution of the story points in the set of all user stories in the project.
In other words, using semantic similarity does not seem effective in discriminating user stories with regard to their story point.
That is a possible explanation for why methods like Deep-SE and TF/IDF-SVM, that rely on semantic similarities between user stories to make a mapping between the groups of similar user stories and their story points, fail to perform better than the Median estimator. Future work might devise and experiment with additional effort drivers extracted from issue reports, in order to achieve more accurate effort estimates.

Moreover, future work can investigate the effect of pre-processing the textual input (i.e., title and description) for example to remove non-natural text like links and code snippets. In fact, the current version of Deep-SE indeed does not perform any pre-processing, however we observed that  6,107 issues in the Tawosi dataset (which accounts for 19.11\% of all issues in this dataset) contains code snippets or stack traces in their description  (see our online appendix \cite{replicationPackage} for a break down of this information per issue type). The presence of such an information may hinders Deep-SE’s learning ability, and could be investigated in future work.

We also observe that Deep-SE is trained by using issues of different types such as Bug, Story, Improvement, etc., (see in Figure \ref{fig_issue_type_pie} the proportion of the different issue types in the Tawosi dataset). Therefore, the structure and content of the issues used as training data in both the original and replication study is heterogeneous, which can hinder the the learning ability of the model. For instance, we observe that there is a statistically significant difference between the length of the description of Bug and Story issue types (two-sided Wilcoxon test $p-value=2.2\mathrm{e}{-16}$ with a medium (0.72) effect size). As mentioned earlier, 19.11\% of the issues in the Tawosi dataset have code snippets or stack traces in their description (see Figure \ref{fig_issue_code_bar} for the number of issue reports including code snippet in their description, along with the total number of issues for each issue type group). 
Although in both studies Deep-SE uses a limited number of tokens (specifically, 100 tokens) from the beginning of the issue context (i.e., title+description) to train and test, it would be interesting to investigate and compare in future work the performance of the model trained on all issues vs. a model trained on one issue type at time (e.g., bug report).

We hope that the results of our study will encourage further work aiming at improving methods for Agile Software Effort Estimation, as well as further prove the importance of replication studies and making replication packages publicly available in order to support reproduction, replication and extension of previous work: \textit{``If I have seen further [than others], it is by standing on the shoulders of giants."}\footnote{Newton, Isaac. "Letter from Sir Isaac Newton to Robert Hooke". Historical Society of Pennsylvania. Retrieved 7 June 2018.}

\section*{Acknowledgments}
Vali  Tawosi, Rebecca Moussa  and  Federica  Sarro are  supported  by the ERC Advanced fellowship grant no. 741278.
We would like to thank the authors of the original papers \cite{deep2018,porru2016,fu2022gpt2sp} for being cooperative, replying to our queries, and supporting this replication.

\bibliographystyle{IEEEtran}
\bibliography{sp-replication}

\begin{IEEEbiography}[{\includegraphics[width=1in,height=1.25in,clip,keepaspectratio]{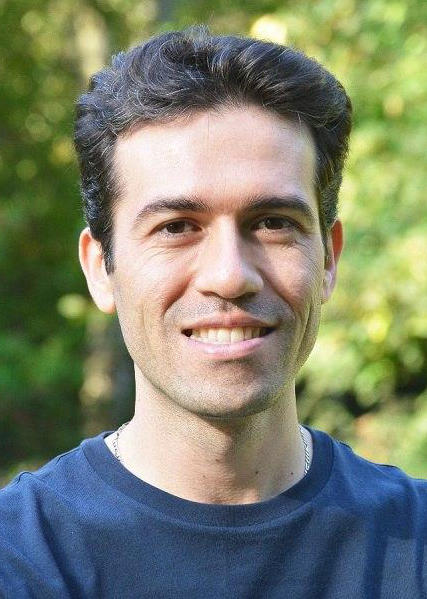}}]{Vali Tawosi} is a PhD student in Software Engineering at University College London, advised by Prof. Federica Sarro and Prof. Mark Harman. His research covers Predictive Analytics for Software Engineering and Search-Based Software Engineering with a focus on software effort estimation and agile methods. Vali is also working as a full-time Researcher at JP Morgan AI Research.
Web page: \url{http://www0.cs.ucl.ac.uk/people/V.Tawosi.html}
\end{IEEEbiography}
\begin{IEEEbiography}[{\includegraphics[width=1in,height=1.25in,clip,keepaspectratio]{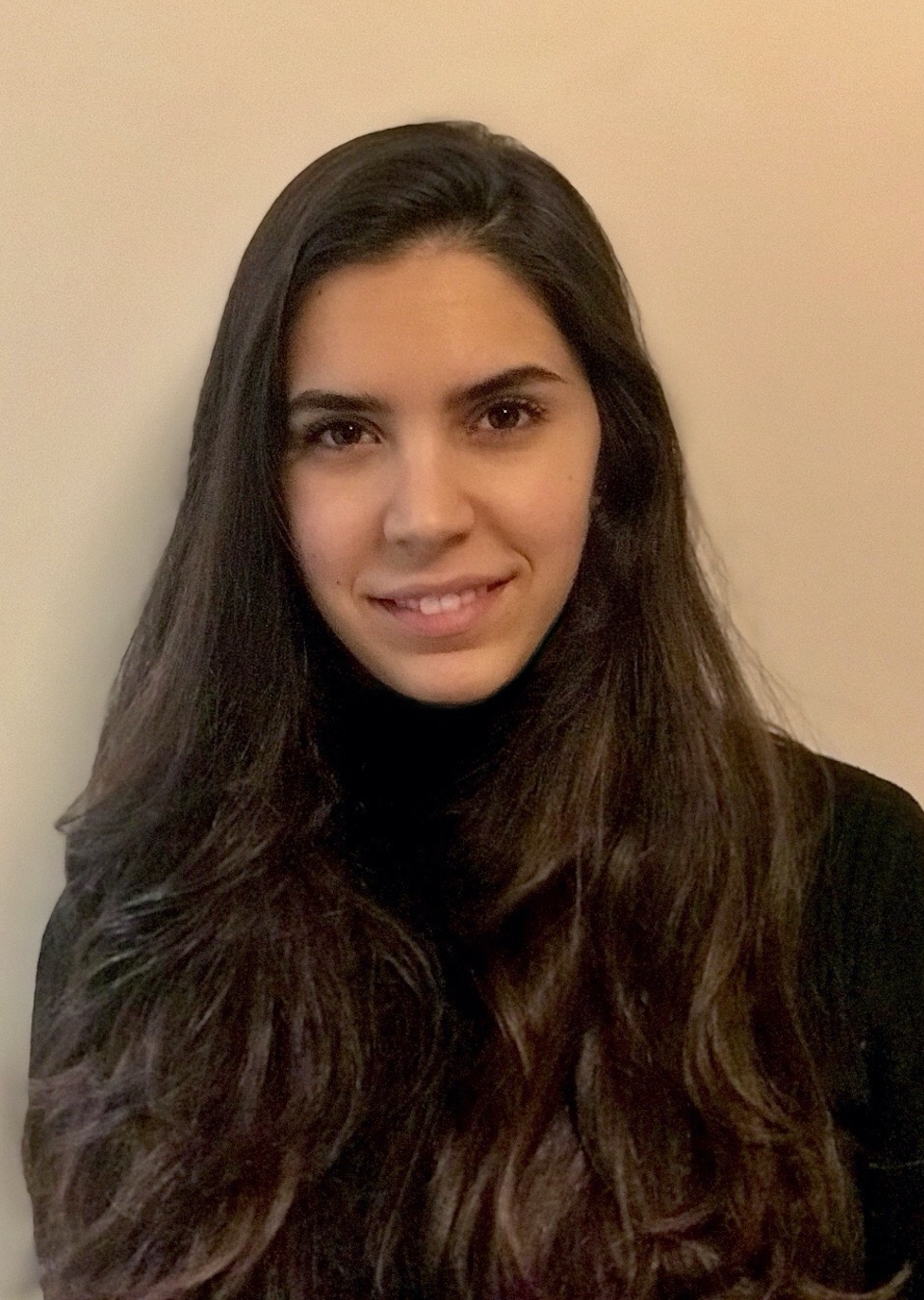}}]{Rebecca Moussa} is a PhD student in Software Engineering at University College London, advised by Prof. Federica Sarro, and Prof. Mark Harman. Her research interests lie in the areas of Predictive Analytics for Software Engineering and Search-Based Software Engineering with a focus on software effort estimation and defect prediction. Web page: \url{http://www0.cs.ucl.ac.uk/people/R.Moussa.html}
\end{IEEEbiography}
\begin{IEEEbiography}[{\includegraphics[width=1in,height=1.25in,clip,keepaspectratio]{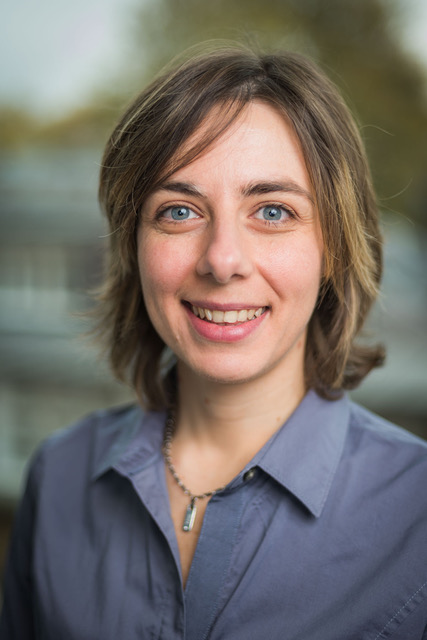}}]{Federica Sarro} is Professor of Software Engineering and Head of the Software Systems Engineering group at University College London.	
Her research aims to push the boundaries for Automated Software Engineering with a focus on Software Analytics, Management, Testing and Optmisation. 
On these topics she has published over 80 papers in peer-reviewed international venues including ICSE, FSE, TSE, TOSEM, EMSE.
Prof. Sarro has received several international awards for her research work including the IEEE TCSE Rising Star Award and the FSE'19 ACM Distinguished Paper Award.
She has been elected and invited to serve on several steering, organisation and programme committees, and programme and editorial boards of well-renowned venues such as ICSE, FSE, ASE, ACM TOSEM, IEEE TSE, IEEE TEVC. Web page: \url{http://www0.cs.ucl.ac.uk/staff/F.Sarro/}
\end{IEEEbiography}

\end{document}